\documentclass[manuscript]{acmart}
\AtBeginDocument{%
  }

\setcopyright{acmlicensed}
\copyrightyear{2026}
\acmYear{2026}
\acmDOI{XXXXXXX.XXXXXXX}

\acmJournal{TOIS}
\acmVolume{Special Issue on Query Performance Prediction Towards Novel Information Retrieval Paradigms}
\acmNumber{x}
\acmArticle{xxx}
\acmMonth{8}

\usepackage{url}
\usepackage{enumitem}
\usepackage{makecell}

\usepackage{caption}
\usepackage{subcaption}

\usepackage{booktabs}
\usepackage{multirow}

\usepackage{pifont}
\newcommand{\cmark}{\ding{51}}%

\usepackage{adjustbox}
\usepackage{tikz}
\usetikzlibrary{positioning}

\newcommand{\para}[1]{\paragraph{\textnormal{\textbf{#1}}}}

\DeclareMathAlphabet{\pazocal}{OMS}{zplm}{m}{n}
\DeclareMathAlphabet{\pazobfcal}{OMS}{cmsy}{b}{n}

\newcommand{\uls}{\begin{itemize}[leftmargin=*]}
\newcommand{\ule}{\end{itemize}}
\newcommand{\ols}{\begin{enumerate}[leftmargin=*]}
\newcommand{\ole}{\end{enumerate}}
\newcommand{\li}{\item}
\usepackage{subcaption}
\usepackage{tcolorbox}
\usepackage{tabularx}
\usepackage{hyperref}

\usepackage[table, x11names]{xcolor}

\usepackage{annotations}

\begin{document}

\title[RAQG-QPP: QPP with Retrieved Query Variants and RAG]{RAQG-QPP: Query Performance Prediction with Retrieved Query Variants and Retrieval Augmented Query Generation}

\thanks{This is a preprint version of an accepted manuscript. The final version will appear in \textit{ACM Transactions on Information Systems}.}


\author{Fangzheng Tian}
\author{Debasis Ganguly}
\author{Craig Macdonald}

\affiliation{%
 \institution{University of Glasgow}
 \country{UK}}

\renewcommand{\shortauthors}{Tian et al.}
\acmArticleType{Review}
\acmCodeLink{https://github.com/borisveytsman/acmart}
\acmDataLink{https://zenodo.org/link}
\acmContributions{FT and DG designed the study; FT conducted the experiments, FT analysed the results, and all authors participated in writing the manuscript.}



\begin{abstract}
Query Performance Prediction (QPP) estimates the retrieval quality of ranking models without the use of any human-assessed relevance judgements, and finds applications in query-specific selective decision making to improve overall retrieval effectiveness. Although unsupervised QPP approaches are effective for lexical retrieval models, they usually perform weaker for neural rankers. Recent work shows that leveraging query variants (QVs), i.e., queries with potentially similar information needs to a given query, can enhance unsupervised QPP accuracy. However, existing QV-based prediction methods rely on query variants generated by term expansion of the input query, which is likely to yield incoherent, hallucinatory and off-topic QVs. 
In this paper, we propose to make use of queries retrieved from a log of past queries as QVs to be subsequently used for QPP. In addition to directly applying retrieved QVs in QPP, we further propose to leverage large language models (LLMs) to generate QVs conditioned on the retrieved QVs, thus mitigating the limitation of relying only on existing queries in a log. Experiments on TREC DL’19 and DL’20 show that QPP enhanced with RAQG outperform the best-performing existing QV-based prediction approach by as much as 30\% on neural ranking models such as MonoT5.
\end{abstract}

\maketitle

\section{Introduction}

Query performance prediction (QPP) is the task of estimating the quality of retrieval results for a given query (referred to as the \textbf{target query}) without relevance judgements. QPP plays an important role in that it can indicate how difficult a query is for a specific retrieval model~\citep{QPP}, and whether retrieval pipeline optimisation is needed~\cite{dlApproachForSeletiveRelevanceF}. As neural ranking models are increasingly applied in Information Retrieval (IR), developing effective QPP methods for those neural models has become an important task. However, unsupervised post-retrieval statistical QPP methods, such as NQC~\cite{NQC} and RSD~\cite{RSD}, have been reported to not perform satisfactorily well for neural rankers~\cite{WRIG}.

The ineffectiveness of unsupervised QPP approaches for neural rankers primarily stems from their reliance on only the target query. As queries are usually short expressions of information needs comprised of a small number of terms, relying on only the input query may not adequately represent the underlying information need~\citep{Oleg_2019}.
To alleviate this problem, incorporating retrieval results from queries related to the target query --- commonly referred to as query variants (\textbf{QV}s) --- can provide additional useful signals to enhance QPP accuracy~\citep{ReferenceBasedQPP}. The effectiveness of leveraging QVs in QPP has been demonstrated by the frameworks proposed by \citet{Oleg_2019} and \citet{WRIG}.

\citet{WRIG} used a relatively simple generative approach based on query expansion (specifically, by means of pseudo-relevance feedback~\citep{NasreenJaleel_RM3}, and skip-gram word vectors~\citep{w2v:mikolov}) to generate QVs in their study. However, the generated QVs often suffer from issues like hallucinations~\citep{Doc2Query--} and topical drift~\citep{LLM-gen-qv}. These term-based approaches for generating query variants
may be limited by restricted vocabularies and may not effectively capture the target query's underlying information need.
We hypothesise that the advent of large language models (LLMs) can be a promising approach for QV generation~\citep{fewShotLearnerLLM} in the sense that LLMs can potentially generate QVs that better encapsulate the target query's information need~\citep{qvIR} and reformulate it in different variations. However, zero-shot outputs from LLMs in QV generation task can be non-realistic and hallucinatory~\cite{detectHallucinations}, in the sense that they may differ from the style of queries that human users can issue to a search engine. To illustrate with an example, when asked to reformulate the query `\textit{lps laws definition}', one of the zero-shot outputs is `\textit{Limited Power of Sale laws: a detailed explanation}', which resembles a declarative title more than an information-seeking query.

We argue that this problem can be solved by appending real-life queries as context to the QV generation process, leading to $k$-shot contextual QV generation. In our proposed method, the queries acting as a context for $k$-shot QV generation are \textbf{retrieved} from a query log containing a large repository of queries submitted by users to a search system. Specifically, for our experiments, we use the MS MARCO training query set, which comprises queries collected from the Bing query log~\citep{MSMARCO-DATASET} with a small number (close to 1 on average) of relevant documents for each query.
We posit that these retrieved \textbf{candidate QVs} from query logs or IR training sets can potentially be useful in grounding an LLM's QV generation process.
Since these query variants potentially contain clues about how information needs evolve within search sessions of different users, they are likely to be more useful for QPP in comparison to the query log agnostic methods, e.g., LLM 0-shot or term-based generation approaches~\citep{WRIG}.

\begin{figure}[t]
\centering
 \includegraphics[width=0.75\columnwidth]{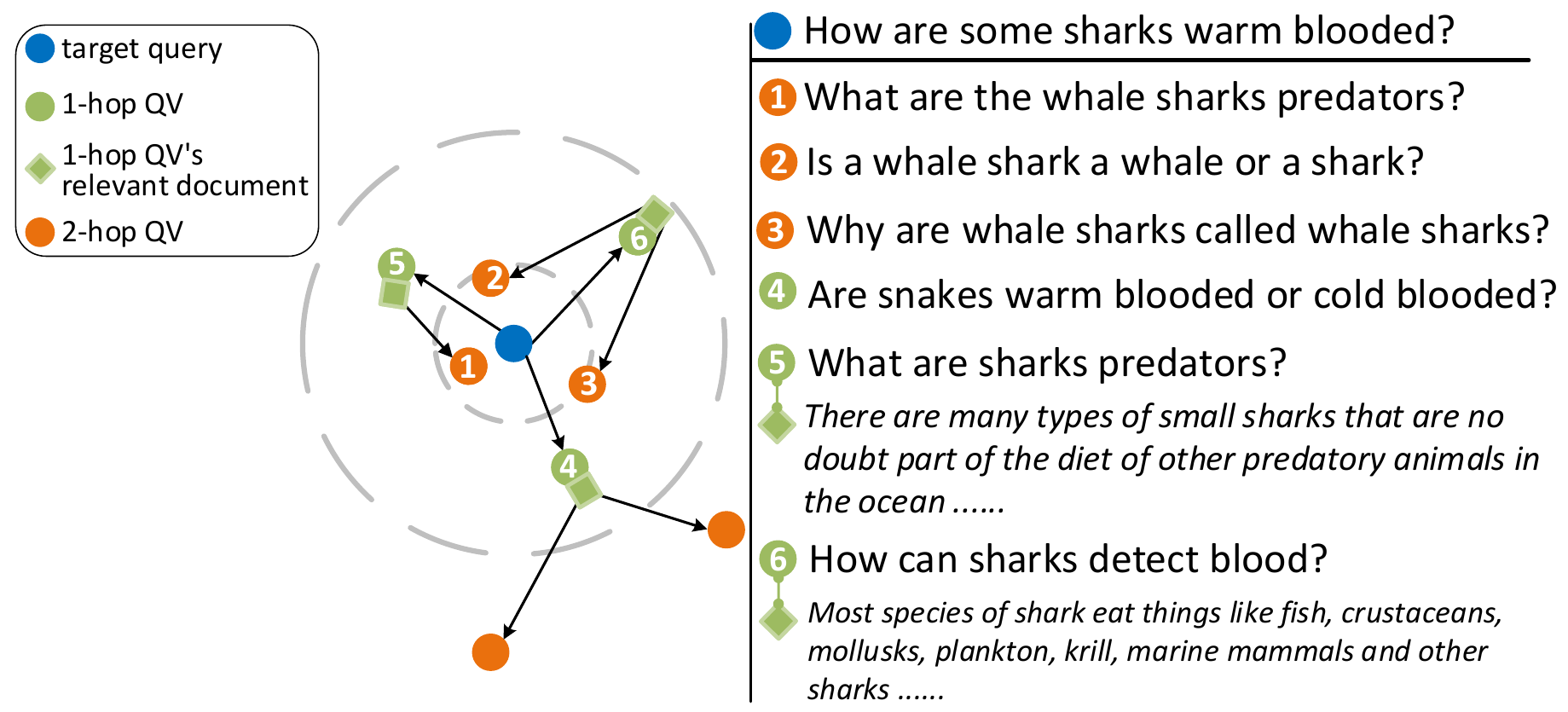}
\caption{A visualisation of the idea behind our proposed QV retrieval method. For a target query (the first line of texts and the blue point in the graph), the first step is to retrieve a list of \textit{1-hop} QVs from an index constructed from the training query set. \textit{1-hop} QVs are represented by the green points. In the next step, the relevant document associated with each \textit{1-hop} candidate (shown as the italicised text) is used as a query to further retrieve a list of \textit{2-hop} QVs from the query index. These \textit{2-hop} QVs are shown in amber. All the \textit{1-hop} and \textit{2-hop} QVs comprise the candidate QV set, which are subsequently ranked into a list (shown as the ordered list on the left) by their topical similarity to the target query (illustrated by their distance to the blue point) to prepare for being used in QPP.
}
\label{fig:neighbourhood}
\end{figure}

Since our proposed QPP method relies on a set of top-retrieved QVs from a query log, an effective QV retrieval method is likely to improve downstream QPP performance. We hypothesise that using only the input query for retrieving similar queries from a collection of queries is limiting in the sense that these input queries are short expressions of information needs. To enrich the target query's representation, we can leverage its associated documents~\citep{GraphReranker}. Therefore, we propose to formulate the relevant documents of the top-retrieved QVs (\textbf{1-hop QVs}) as queries to retrieve more candidate QVs (\textbf{2-hop QVs}), which is conceptually similar to relevance feedback~\citep{RelevanceFeedback}.

Figure~\ref{fig:neighbourhood} illustrates this idea and shows how the initial query neighbourhood, comprised of directly-retrieved 1-hop QVs, is extended by making use of the documents relevant for each of these queries. Taking `\textit{how are some sharks warm blooded?}' as the target query, the top-ranked 1-hop QV in the first-stage retrieval is `\textit{are snakes warm blooded or cold blooded}', which exhibits a significant topic drift from shark to snake, similar to the target query merely in terms of structural pattern. By conducting a second-stage QV retrieval, the retrieved 2-hop QVs, which contain relevant information of the 1-hop QVs, can still be topically related to the target query and act as useful references in QV-based QPP models. For the example in Figure~\ref{fig:neighbourhood}, the 2-hop QV `\textit{what are the whale sharks predators?}' shifts the topic back to sharks from snake.

Moreover, these retrieved QVs can be used as contexts to guide an LLM to \textbf{generate QVs}. Inspired by Retrieval-Augmented Generation (RAG), which enriches the input query with contextual information to improve LLM's performance in downstream tasks, we apply this paradigm to QV generation. Coming back to our example target query in Figure~\ref{fig:neighbourhood}, taking the top-retrieved 2-hop QV as context, LLM can generate `\textit{What are the characteristics of shark species that have endothermic metabolism?}', which represents a different manifestation of an information need largely similar to the target query. We refer to this novel QV generation method as retrieval-augmented query generation (\textbf{RAQG}), and the QPP method based on it as \textbf{RAQG-QPP}. Although LLM-based QV generation has been initially explored by \citet{LLM-gen-qv}, RAQG does not involve human-written backstories, which is more practical in real-life QPP applications.

\begin{figure}[t]
\centering
 \includegraphics[width=0.99\columnwidth]{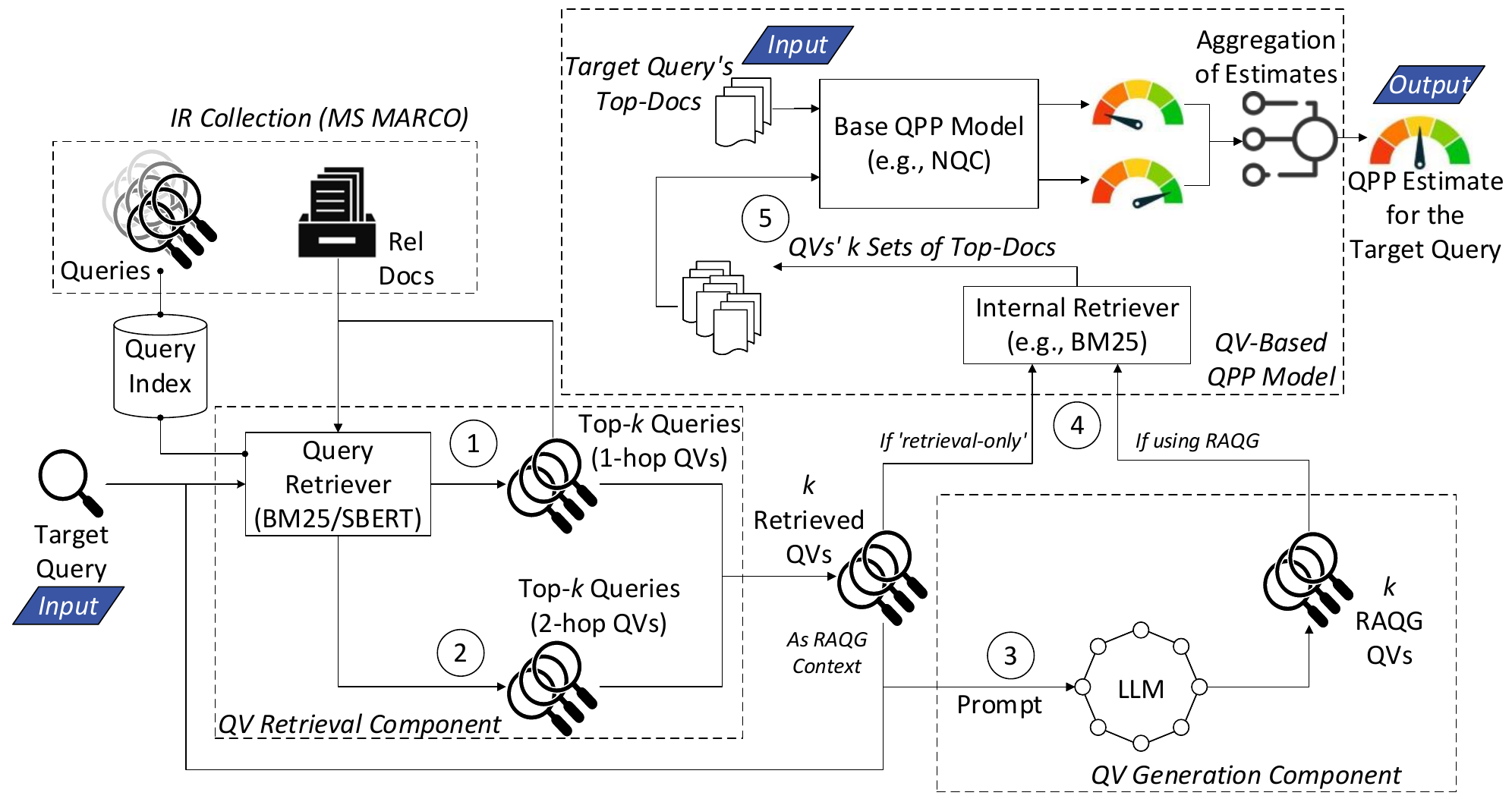}
\caption{The workflow of our proposed RAQG-QPP system, which includes following steps: (1) the input target query is used to retrieve a set of 1-hop query variants (QVs), (2) the relevant documents of 1-hop QVs are then used as augmented queries to retrieve a set of 2-hop QVs, which are then used as context to (3) conditionally generate a set of RAG QVs, which are then used to (4) retrieve a reference list of top-documents by an internal retriever to be used as (5) inputs to a base QPP estimator (e.g., NQC) to obtain the overall estimate.  
\label{fig:flowchart}
}
\end{figure}

The workflow of RAQG-QPP is illustrated in Figure~\ref{fig:flowchart}. After merging and re-ranking the retrieved 1-hop and 2-hop QVs (\textcircled{1} and \textcircled{2} in Figure~\ref{fig:flowchart}), the workflow allows for the following two
configuration options:
(a) the top-retrieved QVs serve as a context to guide the QV generation mechanism
(\textcircled{3} in Figure~\ref{fig:flowchart});
and
(b) the top-retrieved QVs are directly applied for QPP estimation (\textcircled{4} in Figure~\ref{fig:flowchart}).
This design offers a flexible workflow, allowing RAQG to be
configured in various ways according to the effectiveness and efficiency requirements in real-life applications.
As for the QPP module (\textcircled{5} in Figure~\ref{fig:flowchart}), we follow the methodology of \cite{Oleg_2019}, aggregating QPP estimates across the target query and its QVs to produce the final prediction.

In summary, the \textbf{novel contributions} of this paper are two-fold:
\uls
\li We propose a novel method of obtaining useful query variants for the task of QPP, leveraging the known relevant documents from a training query set or query log.
\li In addition to retrieving potentially more useful query variants for QPP, we propose RAQG, a RAG pipeline that leverages these retrieved QVs as context for generating query variants (QVs).
\ule

To facilitate reproducibility and future research on this topic, we release our code at \url{https://github.com/DanielTian97/RAQG_QPP.git}.

In the next section, we review the work related to QV-based QPP and RAG. Sections~\ref{sec:r_qv} and \ref{sec:g_qvs} introduce the methodology of QV retrieval and RAQG, respectively. Section~\ref{sec:experiment_settings} describes the settings of our experiments for validating the effectiveness of RAQG QVs in QPP. Section~\ref{sec:results} answers research questions about RAQG-QPP via analysing the experiment results. Section~\ref{sec:discussions} investigates the mechanism behind the proposed method's improvements in QPP accuracy. Finally, Section~\ref{sec:conc} concludes this paper with directions for future work.

\section{Related Work}\label{s:related_work}

The aim of the proposed QPP framework is to enhance the accuracy of unsupervised QPP methods in neural rankers by leveraging predictions about the target query's QVs. For this reason, we review unsupervised QPP in Section~\ref{ss:unsupervised_qpp} and QV-based QPP methods in Section~\ref{ss:qv_qpp}. Because we applied LLM in Retrieval-Augmented Query Generation (RAQG), in Section~\ref{ss:llm_and_ir}, we review LLM's integration with retrievers.

\subsection{Unsupervised QPP methods}\label{ss:unsupervised_qpp}
QPP methods can be categorised into pre- and post-retrieval ones based on whether the retrieval results are available at prediction time. In \textbf{pre-retrieval QPP} methods, a query's performance is estimated by only using the information from the query and the collection statistics \citep{pre-retrieval-survey}, and thus their prediction objective is either the specificity~\citep{NeuralPreQPP} or difficulty~\citep{embWVQPP} of a query. It can be interpreted as the prediction of the expected retrieval quality, which is not specific to any retrieval models. In contrast, \textbf{post-retrieval QPP} methods are capable of utilising useful signals from the content and the scores of the top-retrieved documents to predict the retrieval quality \citep{Clarity}. The underlying hypothesis common to various post-retrieval predictors is the separability of the top-retrieved documents from the remaining ones: a high separability is a likely indicator of the relevance of the top-retrieved documents~\citep{ImproveQPPbyStandardDeviation}.

To avoid the effect of outliers, some QPP approaches randomly sample subsets of the top-retrieved documents and then aggregate the predictions over each subset, e.g., RSD \citep{RobustStandardDeviationQPP} aggregates the standard deviations~\cite{NQC} of the sampled subsets, whereas UEF \citep{uef_kurland_sigir10} constructs a relevance feedback model from each sampled subset and uses it to estimate the robustness of the retrieval result by computing the relative changes in rankings.

Generally, the effectiveness of unsupervised QPP is largely influenced by retrieval score distribution, especially for neural ranking models where the score is bounded~\citep{WRIG}. In this work, we innovate QV-based QPP with unsupervised base predictors in order to achieve a higher QPP effective for neural rankers.

\subsection{QV-based QPP methods}\label{ss:qv_qpp}
In principle, any statistical estimation method can be improved with the availability of a large number of observation points~\citep{statCompNonDetIRSysUTwoDimVar}. 
In the context of QPP, since a post-retrieval estimator relies on the computation of statistical measures such as the variance of the retrieval score as used by NQC~\citep{NQC}, the estimate for a query can be improved by leveraging information from other queries similar to the target query. These queries with information needs related to the target input query are referred to as query variants (QVs) in the literature \citep{Oleg_2019}.
Usually, a QV-based QPP method first generates QVs using pseudo-relevance feedback (PRF) \citep{NasreenJaleel_RM3} or word2vec \citep{w2v:mikolov} embeddings, as proposed in WRIG~\citep{WRIG,RetrievabilityBasedDocumentSelectionForRFwAutoGenQV}. Each generated QV is then used to retrieve an auxiliary list of documents, which act as \emph{reference lists} for a QPP model~\citep{ReferenceBasedQPP}. A base predictor, e.g., NQC, is then applied to each reference list, and a weighted average is calculated to yield the final prediction.

Apart from QV generation, retrieving QVs from the training set has also been performed by \citet{ContextRichQPP} in a supervised way. They concatenate the retrieved QVs and their performances, which are estimated based on the relevance judgements in the training set, to the target query and its retrieval results. Hence, neighbourhood information is encoded in the target query's expression, which can help the cross-encoder to predict its performance. However, the training queries usually only have shallow annotated relevance judgments, which leads to the concatenated QVs' performances being unreliable. Therefore, their training data might be unavoidably noisy, which can potentially influence QPP accuracy. 

Although QV retrieval has been used by \citet{ContextRichQPP}, they simply retrieve QVs by their cosine similarity with the target query, which may not be effective. There lacks a systematic investigation into the methodology of QV retrieval. In our work, we explore how to retrieve the queries most similar to the target query from a training set via leveraging the relevant documents of the candidate QVs. The QVs retrieved by the proposed method exhibit more usefulness in the QV-based QPP task, and can be used as context to generate more QVs.

\subsection{Synergy between LLM and IR}\label{ss:llm_and_ir}
Large language models (LLMs) have been demonstrated to be useful in a range of tasks related to IR, including ranking~\citep{paradeDemonstrationBasedPassageRanking}, relevance estimation~\citep{oneShotLabellingForAutoRelEst} and QPP~\citep{qpp-llm}. 
In addition, the knowledge stored parametrically in LLMs can be used to understand the underlying information of queries, which can help the IR process. For instance, \citet{query2Doc_EMNLP23} used pseudo-documents generated by LLMs to do query expansion.

Moreover, IR can also be applied to improve the generative process of LLMs. Since the advent of GPT-3~\citep{fewShotLearnerLLM}, LLMs have gained the competence of \textit{few-shot learning}, i.e. learning from a few demonstrations~\citep{iclSurvey}. This paradigm is known as in-context learning (ICL)~\citep{iclPerspective}. If a demonstration dataset is provided, one effective way to seek suitable demonstrations from it for a specific task is to retrieve demonstrations according to their similarity to the target task~\citep{demonSearchPred}.
The retrieved contexts for LLM are not restricted to labelled examples; they can also be unlabelled data, such as the retrieved documents in the context for Retrieval-Augmented Generation (RAG)~\citep{ragInKnowledgeIntensiveNLP}. In RAG, for a specific downstream task, knowledge stored in external datasets is retrieved and passed to an LLM to provide additional knowledge~\citep{ragReview} and alleviate hallucinations in the answer~\citep{FLARE}.

To the best of our knowledge, a RAG-based methodology, i.e., providing retrieved queries as context for generating queries, has not been applied to obtain QVs.
Our proposed method RAQG is able to exploit both directions of this synergic relationship in that
(a) we exploit the semantic ability of an LLM to understand the underlying information need of the target query and to generate QVs based on it, and (b) we provide retrieved query examples as contexts for an LLM to mitigate the potential hallucinations in QV generation. 

In the following two sections (Sections \ref{sec:r_qv} and \ref{sec:g_qvs}), we introduce the details of our proposed RAQG QV generation model.

\section{Methodology of QV Retrieval}\label{sec:r_qv}

We first introduce the unsupervised QV-based QPP framework that is used in the proposed methodology in Section~\ref{ss:aggregate_qvs}. We then introduce our proposed QV retrieval methodology in Section~\ref{ss:retrieve_queries}.

\subsection{Smoothing Estimates from Query Variants}\label{ss:aggregate_qvs}

A post-retrieval QPP model is a function $\phi$ which takes a target query $Q$ and a list of documents $L_M(Q)$ retrieved by executing $Q$ on a \textit{target retrieval model} $M$. The QPP model $\phi$ outputs a score to indicate the quality of the list $L_M(Q)$. Formally speaking, $\phi: L_M(Q) \mapsto \mathbb{R}$. The higher the QPP estimate, the higher the likelihood that the retrieved documents are relevant to the query. 

In QV-based QPP models, the predictor function $\phi$ depends not only on a query and a list of retrieved documents but also on a set of queries that, according to the definition of QV, contain similar information needs as the target query $Q$. Formally speaking, $\phi^+: L_M(Q), \pazocal{E}(Q) \mapsto \mathbb{R}$, where $\pazocal{E}(Q)$ denotes a set of variants of $Q$. 

In this paper, we develop our QPP framework based on the QV-based QPP method proposed by \citet{Oleg_2019}, which aggregates the QPP estimates for $Q$ and its QVs by linear combination. Formally, the predictor in \citep{Oleg_2019} is defined as
\begin{equation}
\phi^+_{M}(L_M(Q), \pazocal{E}(Q)) =
(1-\lambda)\phi(L_M(Q)) +
\lambda\!\!\!\!\!
\sum_{Q' \in \pazocal{E}(Q)}
\frac{
\sigma(Q, Q') \phi(L_{M}(Q'))
}{\overline{\sigma}(Q)
},
\label{eq:basic-knn-qpp}
\end{equation}
\noindent where $\sigma$ is a function measuring the similarity of information needs between a pair of queries $\sigma(Q, Q')$, and $\overline{\sigma}(Q) = \sum_{Q' \in \pazocal{E}(Q)}\sigma(Q, Q')$, which is the average over all the QVs in $\pazocal{E}(Q)$.

Equation~\eqref{eq:basic-knn-qpp} linearly combines the QPP estimates for the target query and the QVs (the same base predictor $\phi$ is used for both). The \textit{relative importance} of the QPP estimation for QVs is controlled by the parameter $\lambda \in [0,1]$. This estimation is a weighted average across all the QVs in $\pazocal{E}$, where the similarity measure between a pair of queries $\sigma(Q, Q')$ is used to determine the relative contribution (weight) of each QV $Q'$.
Following~\citet{Oleg_2019}, we use the rank-biased overlap (RBO) \citep{rbo} as $\sigma$. Specifically, RBO measures the similarity between the top-ranked documents of $Q$ and $Q'$, which reflects the similarity of the information needs between them. Taking RBO similarity as the weight for each QV allows a QPP estimator to assign a higher importance to the contribution from a QV that contains similar information to the target query.

In Equation~\eqref{eq:basic-knn-qpp}, the base predictor $\phi$ yields the QPP estimates for QVs based on their retrieval results obtained by the same retriever $M$ as the target query. As shown in Figure~\ref{fig:flowchart}, our proposed QPP framework generalises Equation~\eqref{eq:basic-knn-qpp} by setting an \textit{internal retriever}. Therefore, the JM-based framework of \citet{Oleg_2019} is modified to: 
\begin{equation}
\phi^+_{M'}(L_M(Q), \pazocal{E}(Q)) =
(1-\lambda)\phi(L_M(Q)) +
\lambda\!\!\!\!\!
\sum_{Q' \in \pazocal{E}(Q)}
\frac{
\sigma(Q, Q') \phi(L_{M'}(Q'))
}{\overline{\sigma}(Q)
},
\label{eq:generic-knn-qpp}
\end{equation}
\noindent where $M'$ is an internal retriever in the QPP model. This setting allows the separation of the internal retriever $M'$ from the target retriever $M$. Therefore, when $M$ is a neural ranker, we can still employ an efficient lexical retriever as $M'$. Moreover, considering the base predictor $\phi$ (e.g. NQC) can be ineffective for neural retrievers~\citep{WRIG}, this separation can add reliability to the QPP estimation for the target query. In our main experiments, we set $M' = BM25$. In Section~\ref{sec:discussions}, we compare the QPP accuracy of keeping $M' = BM25$ with setting the internal retriever to be the same as the target one, i.e., $M' = M$.

\subsection{Retrieving the Query Variants}\label{ss:retrieve_queries}

In our proposed QV-based QPP method, the queries corresponding to the candidate set $\pazocal{E}(Q)$ are retrieved from a training set by the `Query Retriever' component as shown in Figure~\ref{fig:flowchart}. The key motivation for using \textit{retrieved QVs} is that these queries are formulated by real users and, thereby, are more likely to represent real-life information needs. Moreover, since these queries (specifically, obtained from the MS MARCO training data \cite{MSMARCO-DATASET}) are collected from real user queries (Bing query log), they are less likely to be hallucinatory than the ones obtained by existing QV generation methods.

\subsubsection{1-hop QV retrieval}

1-hop QV-based QPP approach is the first variant of our proposed QPP model. For this method, the set $\pazocal{E}(Q)$ of Equation \eqref{eq:generic-knn-qpp} corresponds to a set of top-$k$ queries with the most similar information needs as the target query retrieved from a collection of queries $\pazocal{Q}$. This makes $\pazocal{E}(Q)$ depend on a similarity function, which, given a target query $Q$, is then used to compute its $k$ nearest neighbours. This is formally denoted as
\begin{equation}
\pazocal{E}^{1}(Q) \equiv \pazocal{N}_k(Q, \sigma) \subset \pazocal{Q}, \label{eq:topk_queries}  
\end{equation}
where $\pazocal{N}_k$ denotes a neighbourhood of size $k$, and $\sigma$ denotes a similarity function. In practice, we employ the rank-biased overlap (RBO) as the similarity function ($\sigma$) between queries, as suggested by \cite{Oleg_2019,WRIG}. 

As the RBO similarities are the rank-biased overlap between the retrieval results of two queries, it is not possible to directly obtain from an index (sparse or dense) a top-$k$ list sorted by this similarity measure without first obtaining a candidate set.
As such, to obtain the top-$k$ nearest neighbouring queries (as depicted in Figure~\ref{fig:neighbourhood}), we apply a retrieval-re-rank pipeline. In the first stage, we retrieve candidate QVs of $n (>k)$ top-retrieved queries from an indexed query collection $\pazocal{Q}$. Then we re-rank this list according to each candidate's RBO with respect to the target query, and extract the top-$k$ QVs with the highest RBO scores. This set forms the set of 1-hop QVs. Specific choices for retrieving the 1-hop candidates include a BM25 sparse index and a dense index of SBERT \citep{SBERT} embeddings.

\subsubsection{2-hop QV retrieval}

A limitation of retrieving 1-hop QVs, i.e., $\pazocal{N}_k(Q,\sigma)$ of Equation~\eqref{eq:topk_queries}, is that the queries most similar to the input query $Q$ may not be easily retrieved by only using information of $Q$ itself. A likely reason for not being able to retrieve high-quality QVs in the top ranks is the inherent ambiguity in representing an information need by means of a small number of keywords. To alleviate this limitation, we leverage the available relevant documents for each 1-hop query, expanding the candidate QV set with 2-hop QVs, which potentially offers a higher chance of retrieving more effective QVs. 

Let us denote the set of relevant documents for each query $Q' \in \pazocal{Q}$ by $R(Q')$. We argue that the relevant documents for a 1-hop QV $Q'$ is a better representation of its underlying information need than $Q'$ itself, which, in turn, is potentially associated with the target query $Q$. As a concrete realisation of this idea, we (1) construct a pseudo-query $Q'_D$ from each document $D \in R(Q')$, and then (2) search similar queries for $Q'_D$ on the index of queries $\pazocal{Q}$ to obtain a potentially more effective list of 2-hop queries $\pazocal{E}(Q'_D)$.

The idea is visually illustrated in \textcircled{2} of Figure~\ref{fig:flowchart}, where the set of relevant documents for each 1-hop QV is used to retrieve 2-hop QVs. Specifically, we use the relevant documents associated with each 1-hop query variant $Q'$ as surrogate queries to retrieve similar queries from the training query set. This process yields a set of 2-hop query variants that are indirectly connected to the original query via shared relevance signals provided by the 1-hop query variants. 
Formally,
\begin{equation}
\pazocal{E}^{2}(Q) = \bigcup_{Q' \in \pazocal{N}_k(Q,\sigma)} \bigcup_{D \in R(Q')} \pazocal{N}_k(Q_D,\sigma) \label{eq:2hopqueries}, 
\end{equation}
where $\pazocal{E}^{2}(Q)$ denotes the set of 2-hop QVs.

As a next step, we then merge the $\pazocal{E}^{1}(Q)$ and $\pazocal{E}^{2}(Q)$ as the final candidate QV set, from which we select the QVs for QPP:
\begin{equation}
 \pazocal{E}^{+}(Q) = \pazocal{E}^{1}(Q) \cup \pazocal{E}^{2}(Q) \label{eq:final_candidate_set}, 
\end{equation}
where $\pazocal{E}^{+}(Q)$ is the merged QV set containing both 1-hop and 2-hop retrieved QVs of $Q$. Similar to the 1-hop approach, we find the final nearest-$k$ QVs by computing the RBO scores across the QVs in  $\pazocal{E}^{+}(Q)$ and selecting the QVs with the highest-$k$ RBO scores. These top-$k$ QVs can not only be directly used to improve QV-based QPP but can also serve as context QV generation, as discussed in the next section.

\section{From QV Retrieval to Retrieval Augmented Query Generation} \label{sec:g_qvs}

While using retrieved queries from a training query set for QPP can potentially mitigate hallucinations in QVs, on the flip side, this has two major limitations. First, the available QVs for each query are limited to a small set of potentially useful candidates. Second, depending only on retrieved queries (i.e., relying on an offline query collection) may not yield queries with adequately high similarities
to every target query. To mitigate these limitations, we leverage the generative capabilities of large language models (LLMs) and propose \textit{Retrieval-Augmented Query Generation} (RAQG). 
In this section, we first discuss the existing QV generation methodology that is based on semantic expansions (Section~\ref{ss:basic_g_qvs}). Subsequently, in Section~\ref{ss:raqg_metho}, we introduce our proposed RAQG methodology.

\subsection{Query Variants Generated by Semantic Expansions}\label{ss:basic_g_qvs}

The conceptual foundation for query variants (QVs) generation is semantic expansion~\citep{RetrievabilityBasedDocumentSelectionForRFwAutoGenQV,DBLP:conf/cikm/RoyGMJ16}, a technique that enriches the target query with the terms that are associated with it~\citep{queryExpansion_DFR,DBLP:conf/sigir/GangulyRMJ15}. The terms in the generated QVs are selected from a vocabulary according to their association with the query terms. The association between a pair of terms can be estimated based on the co-occurrences between terms in a local or global way~\cite{DBLP:conf/naacl/SenGJ19}. The local occurrences are calculated using a relevance model derived from the top-ranked documents retrieved for the target query. These documents are considered as the Pseudo-Relevance Feedback (PRF)~\citep{PRF}. One widely-used relevance model in QV generation is RM3~\citep{NasreenJaleel_RM3,DBLP:conf/cikm/RoyGMJ16}. In contrast, global co-occurrences are calculated independently of the retrieved documents, thereby extending the vocabulary for QV generation beyond the terms present in the top-ranked results~\cite{DBLP:conf/sigir/GangulyRMJ15}. An effective approach to estimate global co-occurrences involves word embedding models such as Word2Vec~\citep{w2v:mikolov,DBLP:conf/cikm/RoyGBBM18}. These models represent words as static high-dimensional vectors, and the association between two words is quantified by calculating the cosine similarity between their embedding vectors.

In addition to the problem of hallucinations, which is accompanied by generative processes, the above QV-generation methods also exhibit deficiencies rooted in their respective mechanisms. The methods based on local co-occurrences are constrained by the vocabulary from the words present in the top-retrieved documents, which can limit the diversity of generated QVs. Conversely, the methods that estimate global co-occurrences rely on word embedding models, which produce static representations of words. These representations are context-agnostic, meaning that they cannot capture a word's contextual meaning within the target query. Consequently, these methods may fail to accurately parse out the underlying information needs of the target queries, resulting in the generation of QVs with severely drifted information needs.

\subsection{Retrieval-Augmented Query Generation}\label{ss:raqg_metho}
LLMs - the generator backbone of a RAG pipeline - can be utilised as tools for the semantic expansion of a target query~\citep{query2Doc_EMNLP23}. Therefore, they have the potential for generating query variants (QVs) while addressing the deficiencies of the aforementioned methods that rely on estimated co-occurrence between words. First, it generates QVs using a vocabulary that encompasses all the possible tokens, thereby eliminating the constraints imposed by top-ranked retrieval results. Second, by processing words through their contextual embeddings, LLM-based QV generation methods can more effectively capture the underlying information needs of a target query.

\begin{figure}
\centering
 \includegraphics[width=1\columnwidth]{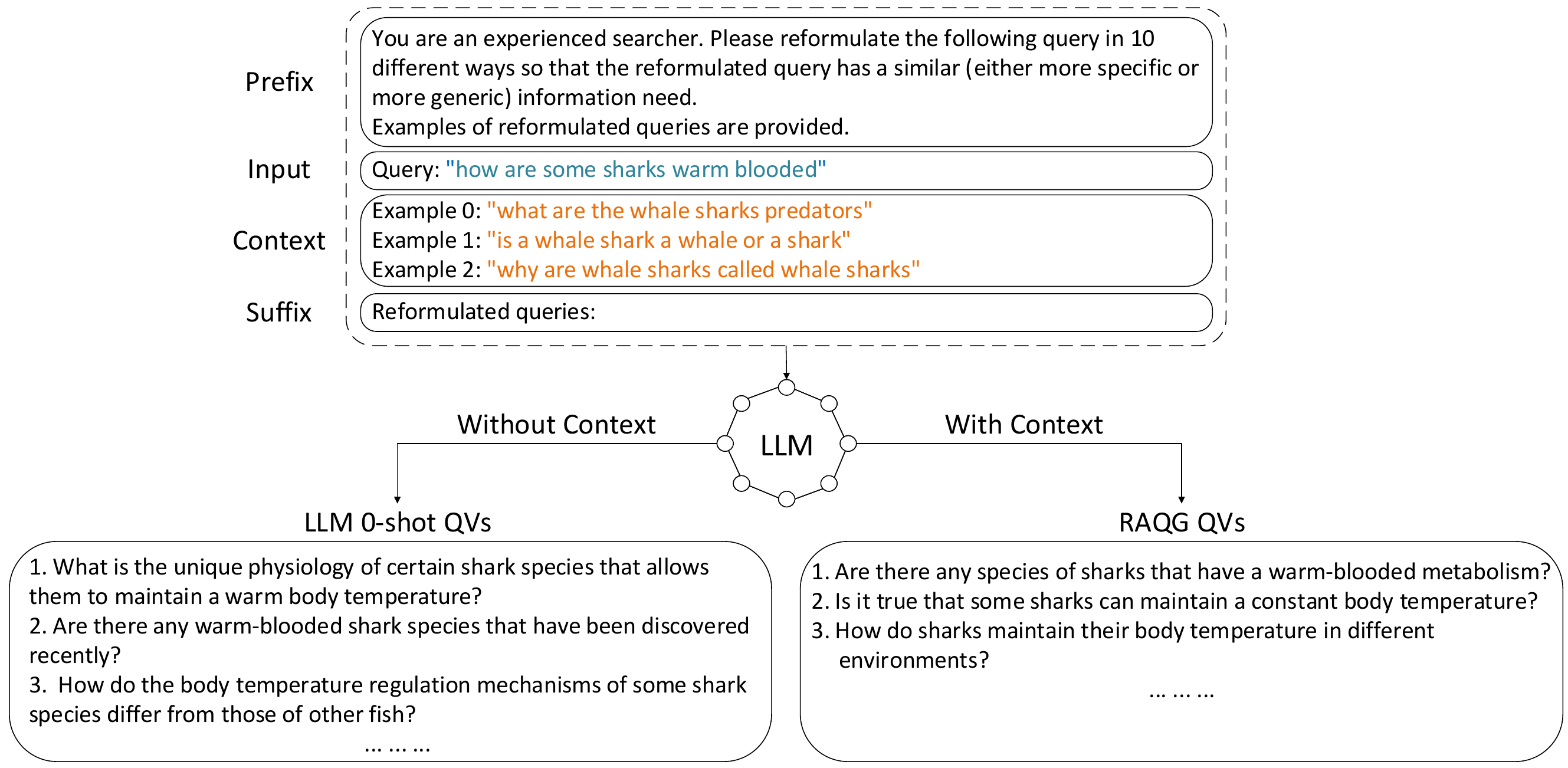}
\caption{A comparison between the QVs generated by Llama3 8B with the non-contextual (0-shot) and the contextual (RAQG) setup for a TREC DL'19 query `\textit{how are some sharks warm blooded}'. The contexts in the shown prompt are the top-3 2-hop neighbouring queries retrieved from the MS MARCO training set, as described in Equation~\eqref{eq:final_candidate_set}. The 0-shot generated queries (left) tend to add more information needs to the target query, e.g., `\textit{Are there any warm-blooded shark species that have been discovered recently?}' and become overly long. In comparison, the RAQG QVs (right), guided by the retrieved queries,  remain concise and better aligned with the target query's underlying information need, closely resembling queries real users might ask.
}
\label{fig:prompt_example}
\end{figure}

\para{Zero-shot LLM-based QV Generation}
In this method, we prompt an LLM to reformulate the target query. The expected output is a list of queries containing information needs that are semantically related to those of the target query. Figure~\ref{fig:prompt_example} shows the prompt template used in our experiments. The figure shows the output obtained with both non-contextual (0-shot) and the contextual generation (RAG) for a sample TREC DL'19 query: `\textit{how are some sharks warm blooded}'.
In particular, as the generation model, we employ the instruction-tuned version of Llama3 8B\footnote{\url{https://huggingface.co/meta-llama/Meta-Llama-3-8B-Instruct}}. The model is quantised for easier deployment and faster inference time, which makes the proposed method more practical in real-time QPP scenarios.

\para{RAG-based QV Generation}
While zero-shot generation is likely to capture the broad information needs of queries, we hypothesise that incorporating the context of the retrieved QVs can serve as an effective controlling factor in the LLM generative process~\citep{ragInKnowledgeIntensiveNLP}. By providing relevant context, the retrieved QVs can potentially guide an LLM's generation process by incorporating queries submitted by real users to a search system. Especially, as the retrieved QVs share similar topics as the target query, they can give hints on how related queries are typically formulated.

In the RAG-based QV generation, we propose to select QVs from the candidate set $\pazocal{E}^+(Q)$, which is merged from 1-hop and 2-hop queries as described in Equation~\eqref{eq:final_candidate_set}, as the context.
The reason for selecting context from $\pazocal{E}^+(Q)$ is that since this set contains both 1-hop and 2-hop queries, the candidate QVs from this set with the highest RBO values are more likely to be topically related to the target query. Hence, they may serve as more useful contexts in the RAG-based QV generation than the ones selected only from 1-hop neighbour queries.

Formally speaking, the QV set of Equation~\eqref{eq:topk_queries} is now defined as
\begin{equation}
\pazocal{E}(Q) = \theta_{\text{LLM}}(Q,
\pazocal{E}^+_p(Q)),
\label{eq:rag}
\end{equation}
where $\theta_{\text{LLM}}$ refers to the frozen decoder parameters of an LLM, and $\pazocal{E}^+_p(Q)$ denotes RAG context, which contains the top-$p$ candidates from the merged set of 1-hop and 2-hop QVs ordered by the RBO metric as shown in Equation~\eqref{eq:final_candidate_set}.
In Equation~\eqref{eq:rag}, $p$ is a hyperparameter controlling the size of the contextual information used for the LLM-based generation. 
We refer to our proposed method of generating query variants using RAG (Equation~\eqref{eq:rag}) as Retrieval-Augmented Query Generation (\textbf{RAQG}).

Returning back to Figure~\ref{fig:prompt_example}, we observe that while both the 0-shot and contextually generated QVs are topically related to the target query, the 0-shot generated QVs tend to express more complex information needs and are generally longer, which is uncommon in real-life user queries~\citep{MSQueryStudy_ShiftOnNeuralRankers}, thus indicating that these may be less useful as reference queries for QPP estimators. In contrast, when retrieved QVs from MS MARCO training set (which, being real user queries, reflect realistic expressions of information needs) are included in the QV-generation prompt, the resulting QVs tend to be concise reformulations of the target query's information need. This observed difference highlights the important role of QV retrieval in our proposed RAQG framework. RBO-based re-ranking of the generated QVs is further able to filter out the queries that exhibit topic drift. When used in the QV-based QPP framework as described in Equation~\eqref{eq:generic-knn-qpp}, RAQG QVs can potentially help yield more accurate QPP estimates.

\section{Experiment Settings}\label{sec:experiment_settings}

We now describe the research questions (Section~\ref{ssec:expsetup:datasets}), evaluation methodology (Section~\ref{ssec:expsetup:eval}), retrieval models (Section~\ref{ssec:expsetup:pipelines}), and the experimented predictors (Section~\ref{experimented_qpps}).

\subsection{Research Questions}\label{ssec:expsetup:datasets}

Our experiments aim to evaluate the effectiveness of our proposed QPP workflow (as illustrated in Figure~\ref{fig:flowchart}) in predicting retrieval quality for two-stage retrieval pipelines with neural reranking. As the foundation of the proposed Retrieval-Augmented Query Generation (RAQG), the query variants (QVs) retrieved from an IR training set should perform better in QV-based QPP than the QVs generated by semantic expansion according to the estimated association between words, such as the QVs generation methods with Pseudo-Relevance Feedback and non-contextual word embeddings, as practised by \citet{WRIG}. Additionally, we seek to determine whether extending the 1-hop QVs (see Equation~\eqref{eq:topk_queries} for details) by retrieving 2-hop QVs based on the relevant documents of the 1-hop QVs (see Equation~\eqref{eq:final_candidate_set} for details) contributes to enhancing the accuracy of QV-based QPP.
Therefore, we formulate the first research question, which is about QPP with retrieved QVs:

\uls
\li \textbf{RQ-1}: \textit{Are retrieved query variants from a training set more effective than those generated based on estimated associations between words}? 
\ule

In addition to the QV retrieval component, the LLM-based QV generation component further distinguishes our proposed RAQG from existing QV generation methods. Therefore, we conduct experiments to evaluate the accuracy of QPP with LLM-generated RAQG-QVs. Since the context of RAQG in our proposed methodology can include either 1-hop or 2-hop retrieved QVs, we compare the QPP accuracy between different RAQG configurations, as depicted in the bottom-right part of Figure~\ref{fig:flowchart}. This comparison may be helpful in verifying whether high-quality retrieved QVs can effectively guide the LLM to generate QVs that can enhance the accuracy of QV-based QPP. Thus, our second research question, which addresses QPP with RAQG QVs, is formulated as:

\uls
\li \textbf{RQ-2}: \textit{Do the QVs generated by our proposed RAQG methodology enhance the accuracy of QV-based QPP?}
\ule

\subsection{Experimental Setup}\label{ssec:expsetup:eval}

\para{Datasets}
We conduct our experiments on the MS MARCO passage collection benchmark~\citep{MSMARCO-DATASET} with target queries from the test sets of TREC DL'19~\citep{TREC-DL-2019-overview} and DL'20~\citep{TREC-DL-2020-overview}. These test sets contain an average of 213 relevant assessments per query, which are constructed through depth pooling and active learning~\citep{activelearning}. For the proposed QV retrieval components, the candidate queries are retrieved from the MS MARCO training set, which consists of over 800K examples of query-relevant document pairs.

\para{Evaluation metrics}
As the main QPP accuracy evaluation metric, we employ Kendall's $\tau$, a standard rank correlation measure. Since different target IR measures can cause significant variations in QPP outcomes~\citep{AnalyseVariantInQPP}, we report our experimental results on two target metrics: AP@100 and nDCG@10. While the former is a metric that combines both precision at top ranks with recall, the latter, in contrast, focuses more on the precision of the top-ranked documents. Following the common practice of evaluating retrieval quality on TREC DL'19 and '20~\citep{TREC-DL-2019-overview, TREC-DL-2020-overview}, documents with a relevance label of 2 or higher are considered relevant in the computation of AP@100.

\para{Train-Test Splits and Hyperparameters}

We evaluate all the QPP methods in our experiments with a 2-fold setup. In the first fold, we train the hyperparameters on DL'19 and apply the optimal settings on DL'20. We switch the roles of the train and the test sets in the second fold, i.e., we find an optimal configuration on DL'20 and apply this setting on DL'19. The common hyperparameters of the QV-based QPP estimators that we optimise over the train folds are: (i) $k$, the number of QVs used in the smoothing estimation, and (ii) $\lambda$, the relative importance of the QV-based QPP smoothing (see Equations \eqref{eq:generic-knn-qpp} for the details about these two parameters).

In addition to $k$ and $\lambda$, for RAQG-based QPP approaches, the hyperparameter $p$ controls the number of retrieved queries that act as context for LLM-based QV generation. To reduce the number of experiments to a tractable limit, we do not tune this parameter on the train split; instead, we report our main results with $p=1$, and then later on analyse the sensitivity of this parameter in combination with $k$.

\subsection{Retrieval Pipelines}\label{ssec:expsetup:pipelines}

Since existing work has highlighted the importance of QV-based QPP models for effective QPP on neural retrievers, we evaluate the accuracy of our proposed QPP approach on several two-stage retrieval pipelines, i.e., a first-stage retriever followed by state-of-the-art neural rankers.

\para{Retrievers}

As first-stage retrievers, we study two commonly used retrievers from the sparse and the dense families:

\uls

\li \textbf{BM25}~\cite{BM25}: A classic IR model based on term weighting and document length normalisation. This lexical model can efficiently select documents that are potentially relevant to the target query.

\li \textbf{TCT-ColBERT}~\citep{TCT-ColBERT}: An efficient end-to-end neural retriever trained by distilling knowledge from ColBERT~\citep{ColBERT}. We use this model to test our QPP models with retrieval pipelines with neural first-stage retrievers.

\ule

\para{Rerankers}

Given the top 100 candidate documents retrieved by either BM25 or TCT-ColBERT, we apply neural reranking models in the second stage. We consider two rerankers representing different architectural paradigms: a cross-encoder model and an LLM-based ranker.

\uls

\li \textbf{MonoT5}~\citep{monot5}~\footnote{Huggingface id {\tt castorini/monot5-base-msmarco}}: MonoT5 is a cross-encoder ranking model, which jointly encodes the query and document together to compute a relevance score. 

\li \textbf{RankLlama}~\cite{finetuningLlamaForRetrieval}~\footnote{Huggingface id {\tt castorini/rankllama-v1-7b-lora-passage}}: RankLlama is a pointwise reranking model fine-tuned from the open-source LLaMA family to assign relevance scores to query–document pairs.

\ule

We evaluate all combinations of the studied retrievers and rerankers. Following PyTerrier~\citep{pyterrier} notation, we use `$\gg$' to denote the cascade operator in a retrieval pipeline. For example, the pipeline that reranks BM25 results using MonoT5 is denoted as BM25$\gg$MonoT5. All retrieval pipelines are implemented using PyTerrier~\citep{pyterrier}. The overall retrieval effectiveness of each pipeline on TREC DL 2019 and 2020 is reported in Table~\ref{table:retrieval_metrics}.

\begin{table}[t]
\centering
\caption{Performance of the experimented retrieval pipelines on TREC-DL-2019 and TREC-DL-2020.}
\small    
\begin{tabular}{@{}l rr rr rr rr@{}}
\toprule
 & \multicolumn{2}{c}{BM25>>MonoT5} & \multicolumn{2}{c}{BM25>>RankLlama} & \multicolumn{2}{c}{TCT>>MonoT5}& \multicolumn{2}{c}{TCT>>RankLlama} \\
 \cmidrule(r){2-3} \cmidrule(r){4-5} \cmidrule(r){6-7} \cmidrule(r){8-9} 
 Dataset & AP@100 & nDCG@10 & AP@100 & nDCG@10 & AP@100 & nDCG@10 & AP@100 & nDCG@10 \\
\midrule 
 TREC-DL-2019 & 0.3582 & 0.6995 & 0.3648 & 0.7182 & 0.3952 & 0.6939 & 0.4267 & 0.7457 \\
 TREC-DL-2020 & 0.4037 & 0.6772 & 0.4273 & 0.6977 & 0.4626 & 0.7017 & 0.5190 & 0.7512 \\
\bottomrule
\end{tabular}
\label{table:retrieval_metrics}
\end{table}

\para{Query Variant (QV) Retrievers}

The QV retriever in our proposed methodology, as shown in Figure~\ref{fig:flowchart}, retrieves queries from the MS MARCO training query set. The criteria for QV retrieval can be based on either lexical or semantic similarities between the training queries and the target query. In our experiments, we test the following query retrievers:

\uls

\li \textbf{BM25}~\citep{BM25}: With a lexical query index, we use this classic lexical retrieval model to measure the lexical similarity between the queries and select the top-ranked queries from the index as the query variants.

\li \textbf{Sentence-BERT} (abbr. \textbf{SBERT})~\citep{SBERT}~\footnote{Huggingface id \texttt{sentence-transformers/all-MiniLM-L6-v2}}: A bi-encoder BERT model based on CLS pooling, pre-trained on the MS MARCO training set triples. We employ it to measure the semantic similarity between two queries, retrieving the queries with the highest cosine similarity to the target queries.

\ule

The candidate sets of top-retrieved QVs obtained by the sparse and dense query retrievers (i.e., BM25 and SBERT, respectively) are then re-scored by their RBO~\citep{rbo} values with the target query. The top-ranked queries, after RBO re-ranking, are then incorporated into the smoothing estimation, as described in Equation~\eqref{eq:generic-knn-qpp}.

\subsection{Baselines and Proposed Methods}\label{experimented_qpps}

We experiment with a range of QPP methods as baselines. These baseline methods are primarily unsupervised base QPP models, including base predictors and existing QV-based predictors. 
To compare our proposed methods with the latest QPP models, we also implement a supervised QPP method \cite{qppbert} with its QV-incorporated variation \cite{ContextRichQPP}.
As a naming convention, we use the \textbf{common prefix} `\textbf{QV}' to name the QV-based unsupervised QPP approaches. All of the QV-based unsupervised methods aggregate a base predictor's QPP estimations for the target query and its QVs through JM smoothing as shown in Equation~\eqref{eq:generic-knn-qpp}.
As QPP baselines, we investigate the following approaches:

\para{Baseline Predictors without QVs}
As baseline QPP models without query variants, we apply the following.
\uls
\li \textbf{NQC}~\citep{NQC}, which estimates the retrieval quality by measuring how skewed the distribution of the retrieval scores at the very top-ranks is.
\li \textbf{UEF}~\citep{uef_kurland_sigir10}, which involves sampling various sub-sequences of the top-ranked documents followed by estimating the robustness of each by computing the perturbations in the rank orders before and after relevance feedback. Following the choice by \citet{WRIG}, we employ NQC as the base estimator of UEF.
\ule

\para{Baseline Predictors with QVs}
As baseline QPP models with query variants, we adopt the existing methods to generate QVs through semantic expansions, as introduced in Section~\ref{ss:basic_g_qvs}. Following the experimental settings in the work of~\citet{WRIG}, we evaluate two QV generation methods based on word associations estimated based on Relevance Language Model (RLM) and Word2Vec (W2V), respectively:

\uls
\li \textbf{QV-RLM}: This method generates QVs by selecting the words that exhibit a high likelihood of association with the terms of the input (target) query. The estimation likelihoods are computed by using the relevance model (RLM) constructed from the top-retrieved documents of the target query. Specifically, we follow the practice of ~\citet{WRIG} and employ RM3~\citep{NasreenJaleel_RM3} as the particular choice for the relevance model.

\li \textbf{QV-W2V}: This method is similar to QV-RLM -- the only difference being that here the association between a document and a query term is estimated by their cosine similarity in an embedding space. Specifically, this embedding space was obtained by employing Word2Vec \citep{w2v:mikolov}, a non-contextualised word embedding model. After constructing a matrix of term similarity values, the QVs are constructed by conducting random walks starting from the terms in the target query~\citep{RetrievabilityBasedDocumentSelectionForRFwAutoGenQV}.

\ule

To maintain comparability of the experiment results, the baseline QV-based QPP models also apply the aggregation method described in Equation~\eqref{eq:generic-knn-qpp}, consistent with our proposed methods.

\para{Supervised Predictors}
To compare our QV-based unsupervised QPP methods with their supervised counterpart, we implement the cross-encoder-based QPP method \textbf{BERTQPP}~\citep{qppbert}. We also experiment with a variant of it, namely \textbf{BERTQPP-QV}~\citep{ContextRichQPP}, which considers QVs and their estimated query performance when encoding the target query. Similar to our 1-hop neighbourhood approach, the BERTQPP-QV method of \citep{ContextRichQPP} uses a candidate set of semantically most similar queries from the MS MARCO set as QVs to improve QPP estimators.

In practice, our proposed Retrieval Augmented QV Generation (RAQG) methodology can be configured in two ways. The first is directly leveraging the retrieved QVs in QPP. The second is to use the contextually generated QVs based on the retrieved QVs in QPP. We evaluate both of them to answer RQ1 and RQ2, respectively.

\para{Experimented Proposed Predictors with Retrieved QVs} 
To name the QPP methods with retrieved QVs, we insert the string `$R$' following the prefix `QV'. Additionally, we set the superscript of `$R$' as `1' or `2' to distinguish whether a 1-hop or 2-hop query neighbourhood is considered in the QV retrieval, as introduced in Section~\ref{ss:retrieve_queries}. According to whether the query retriever is the lexical BM25 or semantic SBERT, we test the following four variants of the proposed method:

\uls
\li \textbf{QV-$R^{1}$-BM25}: This method uses 1-hop query variants only, i.e., only uses the set $\pazocal{E}$ of 1-hop neighbouring queries (see Equation~\eqref{eq:topk_queries}) to smooth the predictor. The 1-hop QVs are retrieved by BM25.

\li \textbf{QV-$R^{2}$-BM25}: Like QV-$R^{1}$-BM25, it also uses BM25 as the query retriever. However, it extends the candidate QV set with 2-hop QV selection as shown in Equation~\eqref{eq:final_candidate_set} and Figure~\ref{fig:flowchart}. The superscript `2' on `R' indicates 2-hop QVs are considered in this method.

\li \textbf{QV-$R^{1}$-SBERT}: This is similar to QV-$R^{1}$-BM25 but employs an embedding-space (semantic) approach to retrieve candidate 1-hop QVs. In particular, we use SBERT as the query retriever.

\li \textbf{QV-$R^{2}$-SBERT}: This method employs the same query retriever as QV-$R^{1}$-SBERT. The difference is that it additionally adds 2-hop neighbours into the candidate QV set.
\ule

\para{Experimented Proposed Predictors with RAQG QVs}

This method uses both the QV retrieval and QV generation components in our proposed RAQG methodology, as described in Section~\ref{sec:g_qvs}. To emphasise the application of LLM-based QV generation, each variant of this method is named with a `$G$' following `QV'. We add a superscript on `$G$' to indicate the number of retrieved QVs used as context in the QV generation (if no context is used, the superscript is `0'). In our experiments, we employ instruction-tuned Llama3~\footnote{https://huggingface.co/meta-llama/Meta-Llama-3-8B-Instruct} as the QV generator. To ensure the reproducibility of the results, all QV generations are conducted with the same random seed and temperature. We test the following predictors based on RAQG QVs:

\uls
\li \textbf{QV-$G^0$}: In the 0-shot setting of QV generation, no example QVs are provided to the LLM. The generation results solely depend on the inherent capability of the LLM in understanding and reformulating queries. This method is an important baseline to assess the effectiveness of using retrieved QVs as context in RAQG.
\li \textbf{QV-$G^p$-\{$C$\}}: In this method, the top $p$ ($p > 0$) relevant retrieved QVs are used as context for Retrieval Augmented QV Generation (RAQG). Those QVs, used as context in the RAQG process, are identical to those used in QV-$\{C\}$, adhering to the convention established for proposed QPP methods that leverage retrieved QVs. For example, QV-$G^3$-$R^2$-SBERT denotes a method in which the top three 2-hop QVs, retrieved by SBERT, serve as the context for the QV generation.
\ule

\section{Results}\label{sec:results}

In this section, we first present the results of the retrieval-only approaches in Section~\ref{ss:ret-only}. 
We then evaluate the performance of QPP approaches with RAQG QVs in Section~\ref{ss:rag-qpp-res}. The sensitivity of the prediction performance towards the hyperparameters, namely $k$, $\lambda$ and $p$ (as described in Section~\ref{ssec:expsetup:eval}) is reported in Section~\ref{ss:sensitivity}.

\subsection{Results of QPP with Retrieved QVs}
\label{ss:ret-only}

We first present the main findings of our approach (relative to the baselines) with a single query variant, i.e., we use $k=1$ in the general QV-QPP framework as described in Equation~\eqref{eq:generic-knn-qpp}. Later, we report the performance of our QPP approach over a range of different values of $k$. 


\begin{table}[t]
\centering
\caption{Comparisons between our proposed QPP method (`Ours') and the baselines (`BL') along with the two supervised methods (`Supervised'). QPP effectiveness is measured by Kendall's $\tau$ with AP@100 and nDCG@10 as the target metrics. Reported Kendall's $\tau$ values are averaged over a 2-fold train-test split on TREC DL'19 and DL'20, with $\lambda$ is tunable while $k$ is fixed to 1, as described in Equation \eqref{eq:generic-knn-qpp}. The best results for each metric with a particular base QPP model are underlined. Comparing the best results obtained with NQC and UEF as the base QPP model, the better ones are bold-faced. A `*' alongside one of our proposed QPP approaches over a target metric indicates that the reported result is statistically significant (by Fisher's $z$-test at 99\% confidence level) over the corresponding base predictor.
}
\small
\begin{adjustbox}{width=0.95\columnwidth}    
\begin{tabular}{@{}lll ll ll ll ll@{}}
\toprule
& & & \multicolumn{2}{c}{$\text{BM25$\gg$MonoT5}$}&
\multicolumn{2}{c}{$\text{BM25$\gg$RankLlama}$}&
\multicolumn{2}{c}{TCT$\gg$MonoT5} & 
\multicolumn{2}{c}{TCT$\gg$RankLlama} \\
\cmidrule(r){4-5} \cmidrule(r){6-7} \cmidrule(r){8-9} \cmidrule(r){10-11}
$\phi$ & Type & Method & AP-$\tau$ & nDCG-$\tau$ & AP-$\tau$ & nDCG-$\tau$ & AP-$\tau$ & nDCG-$\tau$ & AP-$\tau$ & nDCG-$\tau$\\
\midrule 
\multirow{7}{*}{\rotatebox[origin=c]{0}{NQC}} & \multirow{3}{*}{\rotatebox[origin=c]{0}{BL}} 
   & NQC & 0.1673 & 0.0274 & 0.2468 & 0.0978 & 0.1196 & 0.0102 & 0.2537 & 0.0960  \\
 & & QV-W2V & 0.2685 & 0.2041 & 0.2880 & 0.1920 & 0.2060 & 0.0843 & 0.2117 & 0.0498 \\
 & & QV-RLM & 0.3308 & 0.1848 & 0.3038 & 0.1916 & 0.2039 & 0.0565 & 0.2089 & 0.0254 \\
\cmidrule(r){3-11}

 & \multirow{4}{*}{\rotatebox[origin=c]{0}{Ours}} 
   & QV-$R^1$-BM25 & 0.3520 & 0.1918 & 0.3971 & 0.1960 & \underline{0.3013} & \underline{0.1261} & \textbf{\underline{0.3094}} & \textbf{\underline{0.1278}} \\
 & & QV-$R^1$-SBERT & 0.3361 & 0.2000 & 0.3506 & 0.1769 & 0.2619 & 0.0910 & 0.2867 & 0.0927 \\ 
 & & QV-$R^2$-BM25 & 0.3694* & 0.2298 & 0.3907 & 0.1928 & 0.2964 & 0.1235 & 0.3085 & 0.1230 \\ 
 & & QV-$R^2$-SBERT & \textbf{\underline{0.4033}}$^*$ & \textbf{\underline{0.2573}}$^*$ & \textbf{\underline{0.4176}} & \underline{0.2190} & 0.2653 & 0.0818 & 0.2941 & 0.1246 \\

\cmidrule(l){2-11}

\multirow{7}{*}{\rotatebox[origin=c]{0}{UEF}} & \multirow{3}{*}{\rotatebox[origin=c]{0}{BL}} 
   & UEF & 0.1679 & 0.0341 & 0.2495 & 0.1043 & 0.1155 & 0.0006 & 0.2328 & 0.0893 \\ 
 & & QV-W2V & 0.2674 & 0.1969 & 0.2817 & 0.2020 & 0.1134 & 0.0815 & 0.2034 & 0.0424 \\
 & & QV-RLM & 0.3247 & 0.1940 & 0.2945 & 0.1863 & 0.1817 & 0.0512 & 0.1999 & 0.0201 \\
\cmidrule(r){3-11}
 
 & \multirow{4}{*}{\rotatebox[origin=c]{0}{Ours}} 
   & QV-$R^1$-BM25 & 0.3438 & 0.1963 & 0.3763 & 0.1916 & \underline{0.2822} & \underline{0.1265} & \underline{0.3088} & \underline{0.1184} \\
 & & QV-$R^1$-SBERT & 0.3481 & 0.2044 & 0.3440 & 0.1820 & 0.2405 & 0.0833 & 0.2651 & 0.1075 \\
 & & QV-$R^2$-BM25 & 0.3597 & 0.2325 & 0.3735 & 0.1950 & 0.2782 & 0.1179 & 0.2713 & 0.1184 \\ 
 & & QV-$R^2$-SBERT & \underline{0.4026}$^*$ & \underline{0.2540}$^*$ & \underline{0.4143} & \textbf{\underline{0.2229}} & 0.2451 & 0.0795 & 0.2726 & 0.1228 \\

\midrule
\rowcolor{lightgray}
\multicolumn{2}{c}{Supervised} & BERTQPP & 0.2277 & 0.1746 & 0.3118 & 0.1510 & \textbf{0.3194} & \textbf{0.2225} & 0.2656 & 0.0693 \\
\rowcolor{lightgray}
\multicolumn{2}{c}{} & BERTQPP-QV & 0.2432 & 0.1728 & 0.3537 & 0.1146 & 0.2900 & 0.1427 & 0.2894 & 0.0707 \\

\bottomrule
\end{tabular}
\end{adjustbox}

\label{table:ret_qv_main}
\end{table}

The findings from our first set of experiments about QPP with retrieved QVs are reported in Table \ref{table:ret_qv_main}. In this table, the two groups of rows correspond to the two base predictors (denoted as $\phi$) that we experimented with -- specifically, NQC and UEF. Each QV* row is either a baseline (suffix W2V/RLM and also indicated by the type `BL') or a variant of our proposed retrieved-QV-based method (names containing an ``R'' following ``QV'' and also indicated as ``Ours'' in the type column).
Although our proposed QPP framework is unsupervised, we nonetheless include the results of supervised BERTQPP and its variant that leverages QVs (BERTQPP-QV) for the purposes of a complete comparison (these values are, however, greyed out as the comparisons between unsupervised and supervised approaches are not fair).
We evaluate QPP for the target retrievers ($M$) BM25$\gg$MonoT5, BM25$\gg$RankLlama, TCT-ColBERT$\gg$MonoT5 (abbr. TCT$\gg$MonoT5), and TCT-ColBERT$\gg$RankLlama (abbr. TCT$\gg$RankLlama).
From Table \ref{table:ret_qv_main}, we report the following observations pertaining to the first research question (RQ1), i.e., whether leveraging retrieved QVs improves QPP effectiveness.

\para{Observation 1: Retrieved QVs yield higher QPP accuracy compared to the QVs generated by RLM and W2V} In general, the baseline QPP methods with W2V- or RLM-generated QVs are more effective than their corresponding base predictors alone for all the ranking pipelines, e.g., compare the $\tau$ values of NQC (0.1673) vs. QV-RLM (0.3308) for the metric AP@100, i.e., the AP-$\tau$ values. This indeed conforms to the observations reported by~\citet{WRIG}.

Our proposed retrieval-based QV approaches \textit{consistently outperform} the ones generated by either RLM or W2V across all rankers, e.g., compare the QV-RLM correlations with the QV-$R^1$-BM25 ones.
The QV-$R^1$-BM25 result is significantly better than NQC (Fisher's $z$ test at a 99\% confidence interval).
It is also interesting to see that for most rankers, our proposed method of QPP with retrieved QVs (1 and 2 hops) outperforms the supervised approaches as well (with TCT>>MonoT5 being the only exception).

\para{Observation 2: Optimal QV retrieval method depends on the target retrieval pipeline}

It can be observed that with the BM25 first-stage retriever, utilising relevant information from 1-hop query variants to retrieve 2-hop query variants leads to higher QPP accuracy. This trend is consistently observed when comparing 1-hop and 2-hop QPP methods within the same QV retrieval strategy. For example, in the BM25$\gg$MonoT5 pipeline with NQC as the base predictor, the 2-hop method QV-$R^2$-SBERT outperforms the 1-hop method QV-$R^1$-SBERT ($0.4033$ vs. $0.3361$). This improvement in correlation is significant at the 80\% confidence level according to Fisher’s $z$ test.

In contrast, when BM25 is replaced with TCT-ColBERT as the first-stage retriever, a different pattern emerges. Although leveraging 2-hop QVs still improves QPP accuracy over the corresponding 1-hop QVs for SBERT-based QV retrieval, leveraging BM25-retrieved query variants yields stronger overall QPP performance than SBERT-retrieved query variants in these pipelines. As a result, QV-$R^1$-BM25 achieves the highest correlation among the evaluated QPP methods. Notably, because TCT-ColBERT substantially improves the underlying retrieval effectiveness, the highest achieved QPP accuracy is generally lower than that observed in BM25-based pipelines (e.g., correlations of about 0.4 for BM25 versus about 0.3 for TCT-ColBERT). This suggests that accurate QPP becomes more challenging as the retrieval pipeline becomes more effective, and that our proposed QPP methods provide improvements over base unsupervised predictors (e.g., NQC and UEF) consistently in this more challenging setting.

\subsection{Results of RAQG-QPP}
\label{ss:rag-qpp-res}\label{ss:main_ob_RAQG-QV}

The observations in response to RQ1 confirm that the retrieved QVs can largely enhance the QPP accuracy of the base predictors. In this section, we further integrate the retrieved QVs as the context in Retrieval-Augmented QV Generation (RAQG) and assess the RAQG component in our proposed workflow.

Table~\ref{table:pshot-k=1} presents the results of RAQG-QPP (NQC as the base predictor) for $k=1$ and $k=5$. In the tested RAQG-QPP methods, the QVs are contextually generated based on the top-ranked query ($p=1$) yielded by the corresponding QV retrieval method. For comparison, Table~\ref{table:ret_qv_main} also reports the experiment results for: (1) the base predictor NQC, (2) the QPP methods based on retrieved QVs, and (3) the QPP methods based on 0-shot generated QVs.
Analysing the experiment results, we highlight the following observations to answer \textbf{RQ2}, which investigates the effectiveness of applying \textbf{RAQG QVs in QPP}:

\begin{table}[t]
\centering
\caption{Comparisons between QPP with RAQG-QVs and retrieved QVs. The `R' and the `G' columns denote whether a method uses QVs from either `retrieval' or `generation' or both. The number of QVs ($k$) is 1 for the second group of rows and 5 for the third group of rows. The best values in each column for every $k$ are underlined, while the comparatively better values are additionally boldfaced. A `$\dagger$' alongside one of our proposed RAQG QPP approaches over a target metric indicates that the reported result is significantly better than the QPP method based on the QVs that are used as RAQG context in it (Fisher's $z$-test at 90\% confidence level). For example. when $k=5$, QV-$G^1$-$R^2$-BM25 is significantly better than QV-$R^2$-BM25, thus it is noted with a `$\dagger$'. The greyed rows are copied from Table~\ref{table:ret_qv_main} for convenience in comparisons.}
\begin{adjustbox}{width=0.95\columnwidth} 
\small
\begin{tabular}{@{}ll cc ll ll ll ll @{}}
\toprule
& & & & \multicolumn{2}{c}{BM25$\gg$MonoT5} & \multicolumn{2}{c}{$\text{BM25$\gg$RankLlama}$}&
\multicolumn{2}{c}{TCT$\gg$MonoT5} & 
\multicolumn{2}{c}{TCT$\gg$RankLlama} \\
\cmidrule(r){5-6} \cmidrule(r){7-8} \cmidrule(r){9-10} \cmidrule(r){11-12}
$k$ & Method & R & G & AP-$\tau$ & nDCG-$\tau$ & AP-$\tau$ & nDCG-$\tau$ & AP-$\tau$ & nDCG-$\tau$ & AP-$\tau$ & nDCG-$\tau$ \\
\midrule
\rowcolor{lightgray}
0 & NQC &  &  & 0.1673 & 0.0274 & 0.2468 & 0.0978 & 0.1196 & 0.0102 & 0.2537 & 0.0960 \\
\cmidrule{1-12}
\rowcolor{lightgray}
\multirow{10}{*}{1} & QV-$R^1$-BM25 & \cmark &  & 0.3520 & 0.1918 & 0.3971 & 0.1960 & \textbf{\underline{0.3013}} & \textbf{\underline{0.1261}} & 0.3094 & 0.1278 \\
\rowcolor{lightgray}
& QV-$R^1$-SBERT & \cmark &  & 0.3361 & 0.2000 & 0.3506 & 0.1769 & 0.2619 & 0.0910 & 0.2867 & 0.0927 \\
\rowcolor{lightgray}
& QV-$R^2$-BM25 & \cmark &  & 0.3694 & 0.2298 & 0.3907 & 0.1928 & 0.2964 & 0.1235 & 0.3085 & 0.1230 \\
\rowcolor{lightgray}
& QV-$R^2$-SBERT & \cmark &  & 0.4033 & \underline{0.2573} & 0.4176 & 0.2190 & 0.2653 & 0.0818 & 0.2941 & \textbf{\underline{0.1246}} \\
\cmidrule{3-12}
& QV-$G^{0}$ &  & \cmark & 0.3452 & 0.2392 & 0.3714 & 0.2061 & 0.2710 & 0.0761 & \underline{0.3412} & 0.1022 \\
& QV-$G^1$-$R^1$-BM25 & \cmark & \cmark & 0.3592 & 0.2238 & 0.3903 & 0.2394 & 0.2473 & 0.0250 & 0.3121 & 0.0892 \\
& QV-$G^1$-$R^1$-SBERT & \cmark & \cmark & 0.3324 & 0.2021 & 0.3601 & 0.2313 & 0.2160 & 0.0615 & 0.2970 & 0.0554 \\
& QV-$G^1$-$R^2$-BM25 & \cmark & \cmark & 0.3845 & 0.2266 & 0.4213 & \underline{0.2584} & 0.2405 & -0.0075 & 0.3082 & 0.0727 \\
& QV-$G^1$-$R^2$-SBERT & \cmark & \cmark & \underline{0.4286} & 0.2124 & \underline{0.4469} & 0.1988 & 0.2812 & 0.0741 & 0.3292 & 0.0575 \\
\cmidrule{1-12}
\multirow{10}{*}{5} & QV-$R^1$-BM25 & \cmark &  & 0.3231 & 0.1837 & 0.3532 & 0.1744 & 0.2591 & 0.0990 & 0.2788 & 0.0533 \\
& QV-$R^1$-SBERT & \cmark &  & 0.3064 & 0.1983 & 0.3409 & 0.1806 & 0.2588 & 0.1178 & 0.2868 & 0.0890 \\
 & QV-$R^2$-BM25 & \cmark &  & 0.3231 & 0.1837 & 0.3215 & 0.1522 & 0.2226 & 0.0907 & 0.2933 & 0.0470 \\
& QV-$R^2$-SBERT & \cmark &  & 0.3425 & 0.1806 & 0.3319 & 0.0782 & 0.1185 & -0.0053 & 0.2470 & 0.0527 \\
\cmidrule{3-12}
& QV-$G^{0}$ &  & \cmark & 0.4063 & 0.2674 & 0.4255 & 0.2498 & \underline{0.2711} & \underline{0.1101} & 0.3397 & 0.1210 \\
& QV-$G^1$-$R^1$-BM25 & \cmark & \cmark & 0.3963 & 0.2608 & 0.4173 & \textbf{\underline{0.2743}} & 0.2428 & 0.0817 & \textbf{\underline{0.3497}} & \underline{0.1222} \\
& QV-$G^1$-$R^1$-SBERT & \cmark & \cmark & 0.3388 & 0.2060 & 0.3677 & 0.2263 & 0.1411 & 0.0045 & 0.2474 & 0.0415 \\
& QV-$G^1$-$R^2$-BM25 & \cmark & \cmark & 0.4412$\dagger$ & \underline{\textbf{0.2716}} & 0.4420 & 0.2732 & 0.2407 & 0.0691 & 0.3334 & 0.1087 \\
& QV-$G^1$-$R^2$-SBERT & \cmark & \cmark & \underline{\textbf{0.4440}}$\dagger$ & 0.2099 & \textbf{\underline{0.4669}} & 0.2032 & 0.2456 & 0.0002 & 0.3342 & 0.0617 \\

\bottomrule
\end{tabular}

\label{table:pshot-k=1}
\end{adjustbox}
\end{table}

\para{Observation 1: Leveraging LLM-generated QVs improves QPP accuracy}
Compared with the base predictor (NQC) in Table~\ref{table:pshot-k=1}, QPP methods incorporating LLM-generated query variants consistently achieve higher accuracy. This indicates that the proposed RAQG framework enables LLMs to generate query variants that effectively enhance QPP. However, 0-shot generated QVs (QV-$G^0$) alone do not consistently outperform the strongest QPP methods based on retrieved query variants.

For retrieval pipelines that rerank BM25's retrieval results, incorporating retrieved QVs, particularly 2-hop retrieved QVs, as contextual guidance for RAQG can generate query variants that yield higher QPP accuracy than the best-performing retrieved-QV-based methods, especially for AP@100. This advantage becomes more pronounced as more RAQG QVs (i.e., larger number of QVs with increasing values of $k$) are incorporated into QPP estimation. For example, when $k=5$, QV-$G^1$-$R^2$-SBERT achieves a Kendall’s $\tau$ of 0.4669 for predicting AP@100 for BM25$\gg$RankLlama.
In contrast, when TCT-ColBERT is used as the first-stage retriever, the benefit of incorporating retrieved QVs as RAQG context becomes marginal. In these pipelines, the best QPP accuracy achieved by retrieved-QV-based methods and RAQG-based methods is usually similar.

Overall, these results support our hypothesis in Section~\ref{sec:g_qvs} that retrieval-augmented query generation can enhance QV-based QPP, while also indicating that the choice of query variants leveraged for QPP should be adapted to the target retrieval pipeline.

\para{Observation 2: 2-hop retrieved QVs mostly provide better RAQG context than 1-hop retrieved QVs.}

Comparing the results obtained with QV-$G^1$-$R^1$-C and QV-$G^1$-$R^2$-C (C can be either BM25 or SBERT) in Table~\ref{table:pshot-k=1}, we observe that the latter almost consistently achieves higher QPP accuracy. The gains in correlation are higher for AP@100 prediction.
For example, when $k=5$ and target retriever is BM25$\gg$MonoT5, QV-$G^1$-$R^2$-SBERT ($0.4440$) outperforms QV-$G^1$-$R^1$-SBERT ($0.3388$) and QV-$G^1$-$R^1$-BM25 ($0.3963$).
This observation establishes the importance of using high-quality retrieved QVs as context in RAQG.

Moreover, this observation further confirms that incorporating relevance information -- specifically, the fact that the 2-hop QVs are derived using the relevant documents of the 1-hop neighbours -- facilitates the retrieval of higher-quality QVs, which in turn yields improved QPP performance. In this context, QV quality is characterised in two ways: first, by its ability to produce high QPP accuracy when used directly in QV-based QPP (as defined in Equation \eqref{eq:final_candidate_set}); and second, by its effectiveness in guiding the RAQG process to generate useful QVs for QPP (as described in Equation \eqref{sec:g_qvs}).

\subsection{Parameter Sensitivity} \label{ss:sensitivity}

As an unsupervised QPP method, it is essential to examine the parameter sensitivity of our proposed method to ensure its robustness in real-life scenarios. In this section, we conduct a parameter sensitivity analysis of the BM25$\gg$MonoT5 retrieval pipeline, as this setting exhibits the most consistent QPP gains, allowing for a controlled analysis of the parameter variation effects.

\begin{figure}[t]
\centering
\begin{subfigure}[h]{0.32\textwidth}
 \centering
 \includegraphics[width=\textwidth]{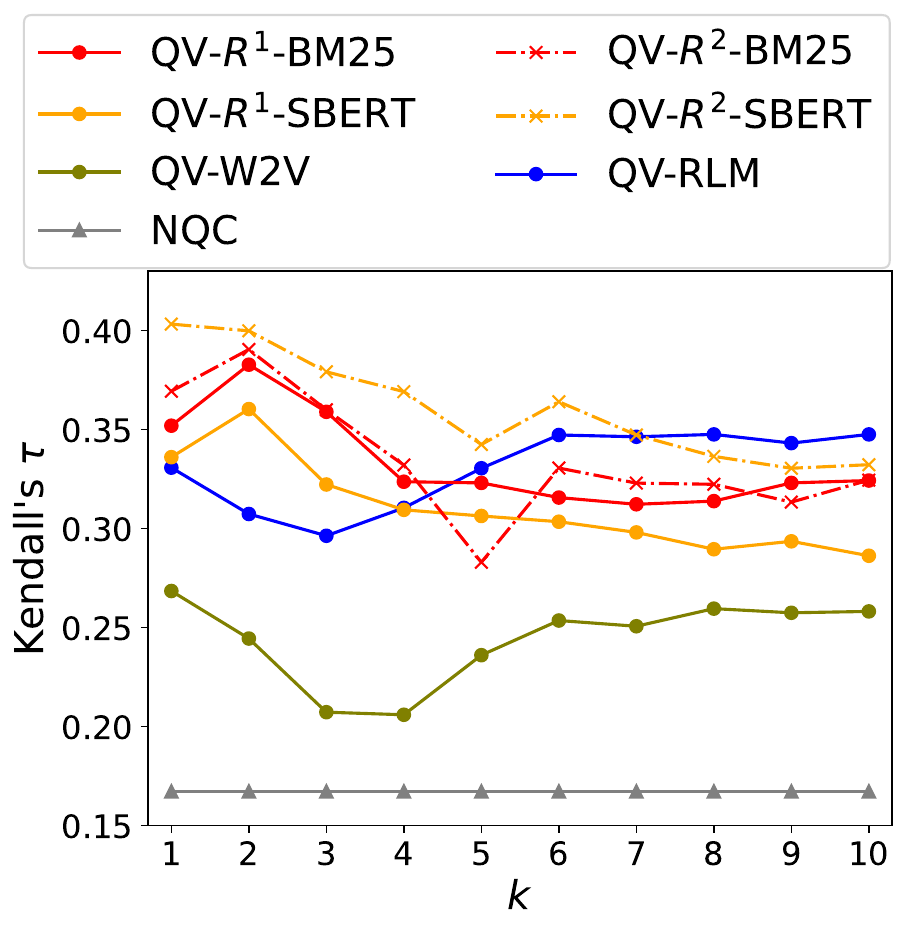}
 \caption{NQC on AP@100}
\end{subfigure}
\begin{subfigure}[h]{0.32\textwidth}
 \centering
 \includegraphics[width=\textwidth]{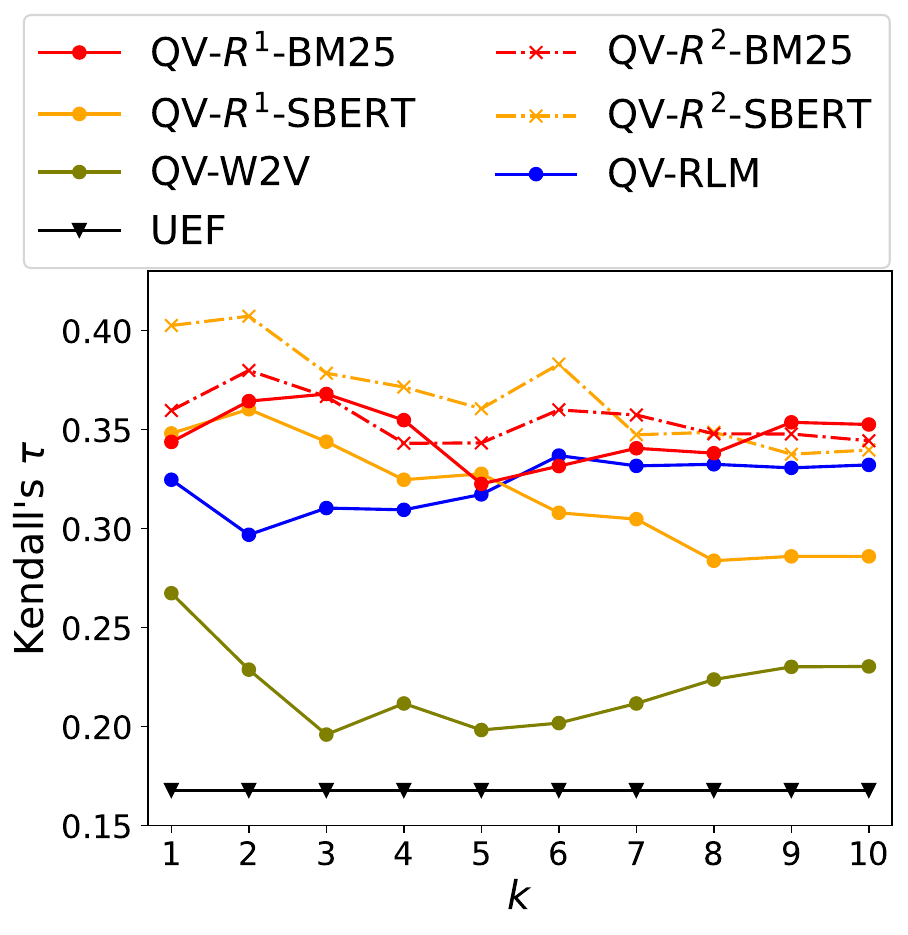}
 \caption{UEF on AP@100}
\end{subfigure}
\vfill
\begin{subfigure}[h]{0.32\textwidth}
 \centering
 \includegraphics[width=\textwidth]{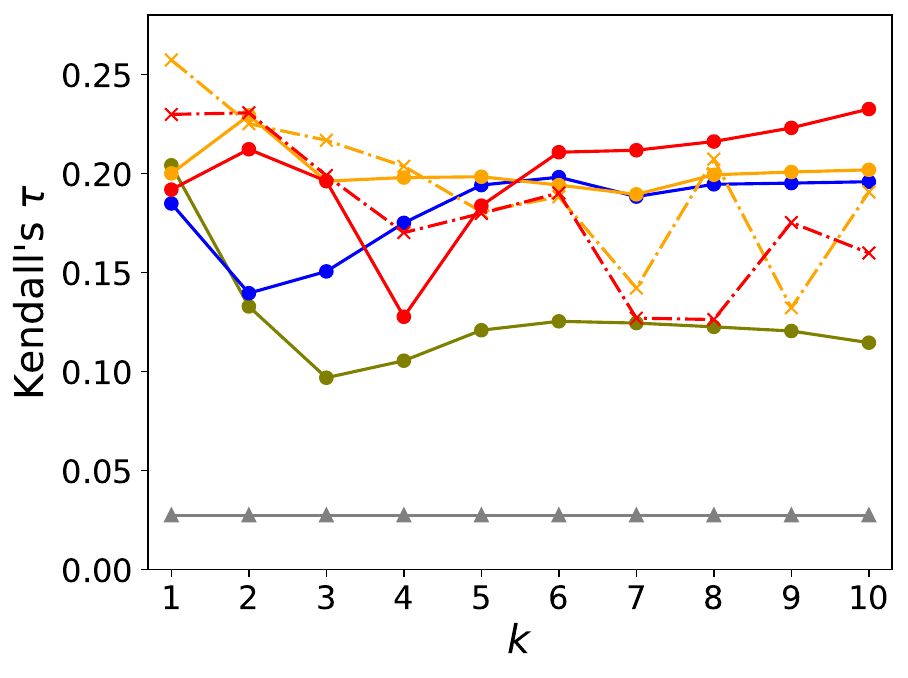}
 \caption{NQC on nDCG@10}
\end{subfigure}
\begin{subfigure}[h]{0.32\textwidth}
 \centering
 \includegraphics[width=\textwidth]{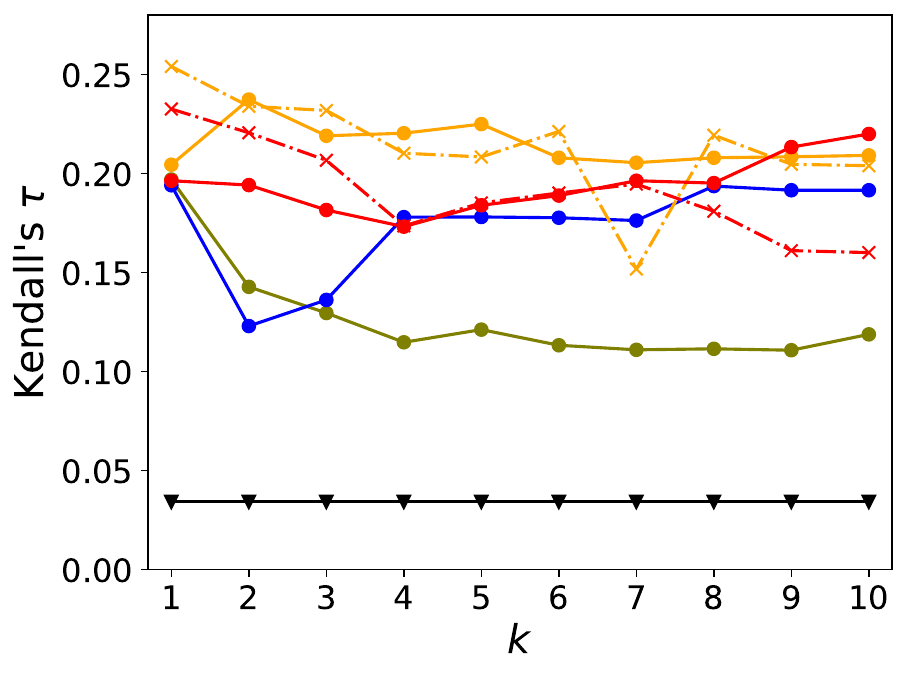}
 \caption{UEF on nDCG@10}
\end{subfigure}
\caption{Performance of baseline and proposed QPP methods across different numbers of query variants used (the parameter $k$). The target metrics are AP@100 for the subplots in the top row and nDCG@10 for the ones on the bottom. The general trend is that QPP accuracy decreases as more QVs are incorporated.}
\label{fig:mt5_full_k}
\end{figure}

\subsubsection{Parameter Sensitivity in QPP Approaches with Retrieved QVs}\label{sss:sensitivity-ret-only}

There are two hyperparameters for QV-based QPP methods: $k$ -- the number of QVs, and $\lambda$ -- the relative importance of the QVs in the final QPP estimate as described in Equation~\eqref{eq:generic-knn-qpp}. We analyse the sensitivity of QPP accuracy to these two parameters for both the baselines and proposed QV-based QPP methods.

\para{Effect of varying $k$ with respect to $\lambda$s in QPP based on retrieved QVs}
Figure~\ref{fig:mt5_full_k} shows the QPP accuracy achieved with incorporating various numbers ($k$) of retrieved QVs ($k\in \{1,\ldots,10\}$). Similar to Table \ref{table:ret_qv_main}, the values of $\lambda$ are optimised on the train splits of the 2-fold setup. The general trend observed in all the configurations of the retrieved QVs is that QPP accuracy decreases with $k$, the number of QVs used. This is expected because, as $k$ increases, QVs with smaller RBO values with the target query (indicative of potential information need drifts) play a role in the QPP estimation.

\begin{figure}[t]
     \centering
     \begin{subfigure}[b]{0.32\textwidth}
         \centering
         \includegraphics[width=\textwidth]{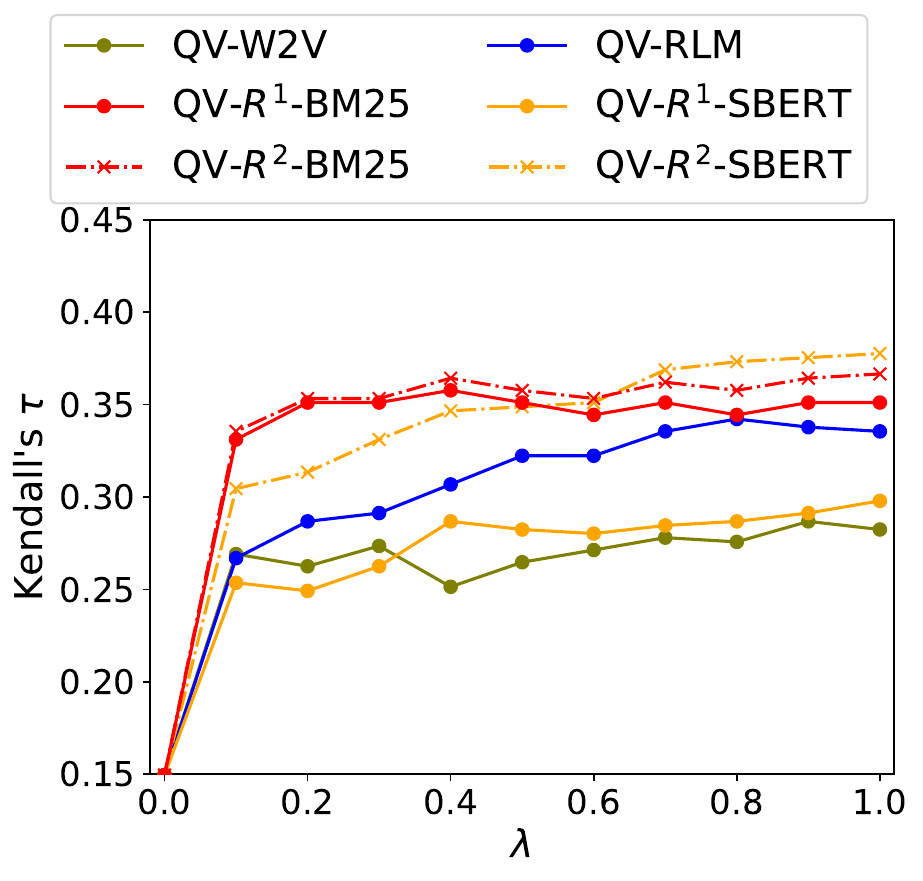}
         \caption{DL'19: $k=1$}
     \end{subfigure}
     \begin{subfigure}[b]{0.32\textwidth}
         \centering
         \includegraphics[width=\textwidth]{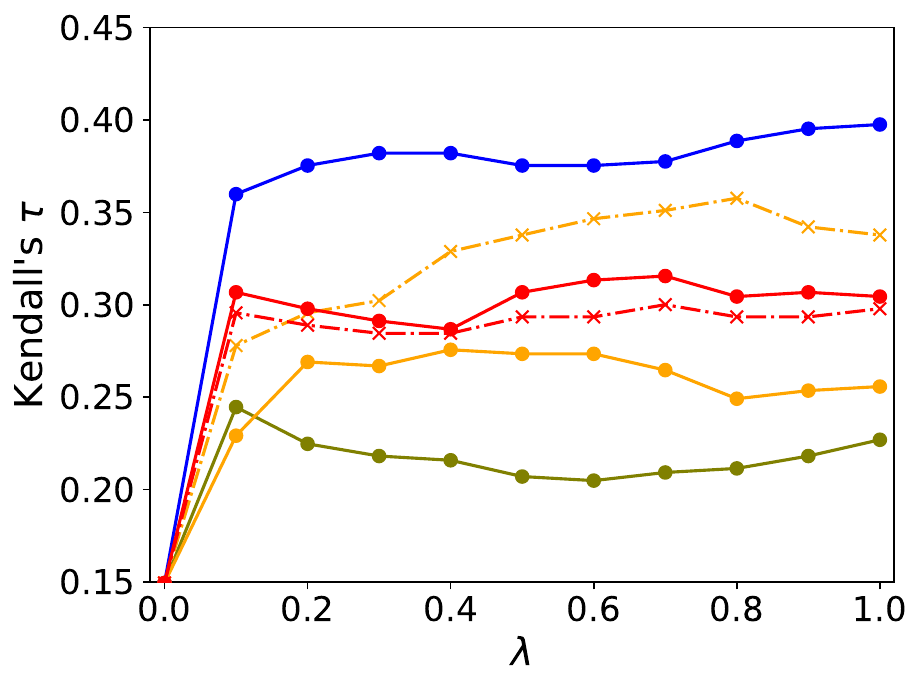}
         \caption{DL'19: $k=3$}
     \end{subfigure}
     \begin{subfigure}[b]{0.32\textwidth}
         \centering
         \includegraphics[width=\textwidth]{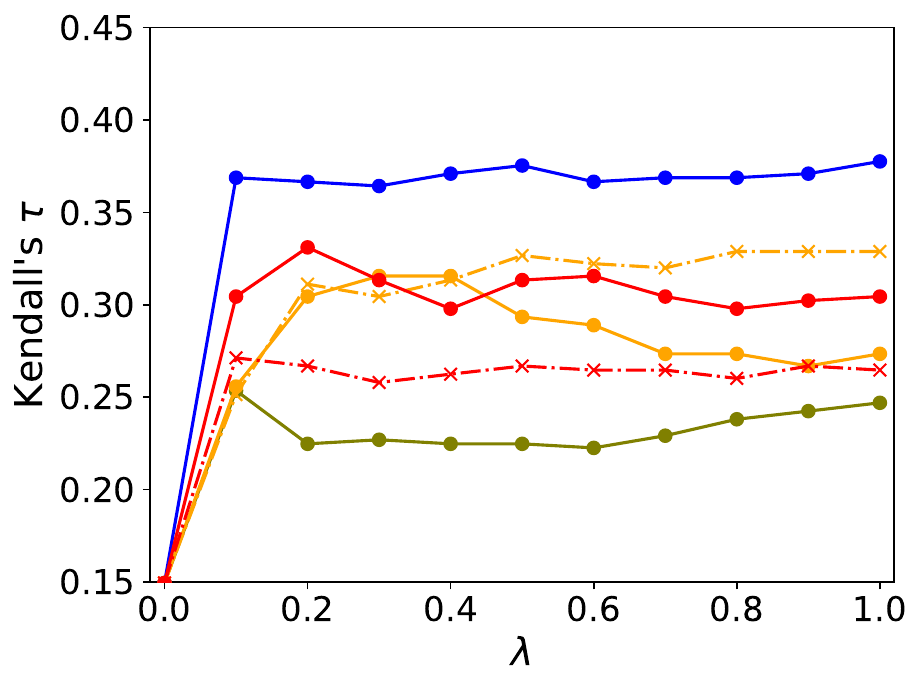}
         \caption{DL'19: $k=5$}
     \end{subfigure}
     \begin{subfigure}[b]{0.32\textwidth}
         \centering
         \includegraphics[width=\textwidth]{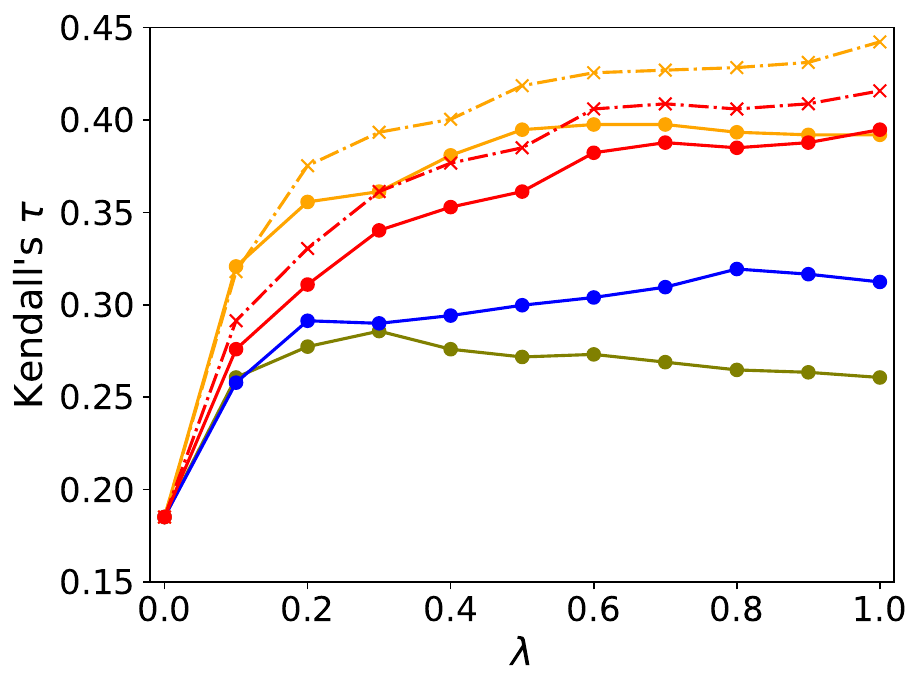}
         \caption{DL'20: $k=1$}
     \end{subfigure}
     \begin{subfigure}[b]{0.32\textwidth}
         \centering
         \includegraphics[width=\textwidth]{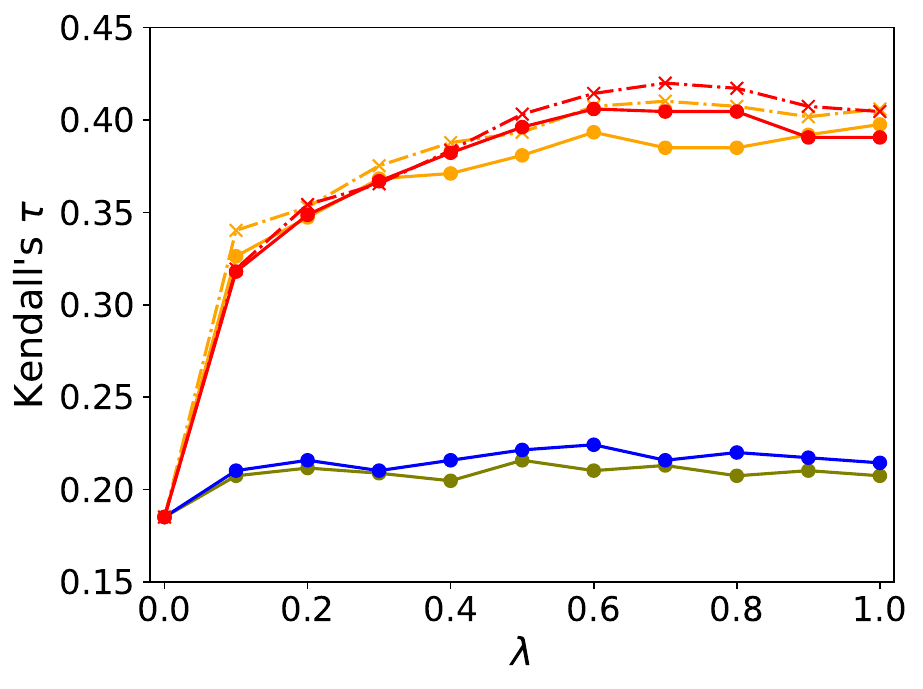}
         \caption{DL'20: $k=3$}
     \end{subfigure}
     \begin{subfigure}[b]{0.32\textwidth}
         \centering
         \includegraphics[width=\textwidth]{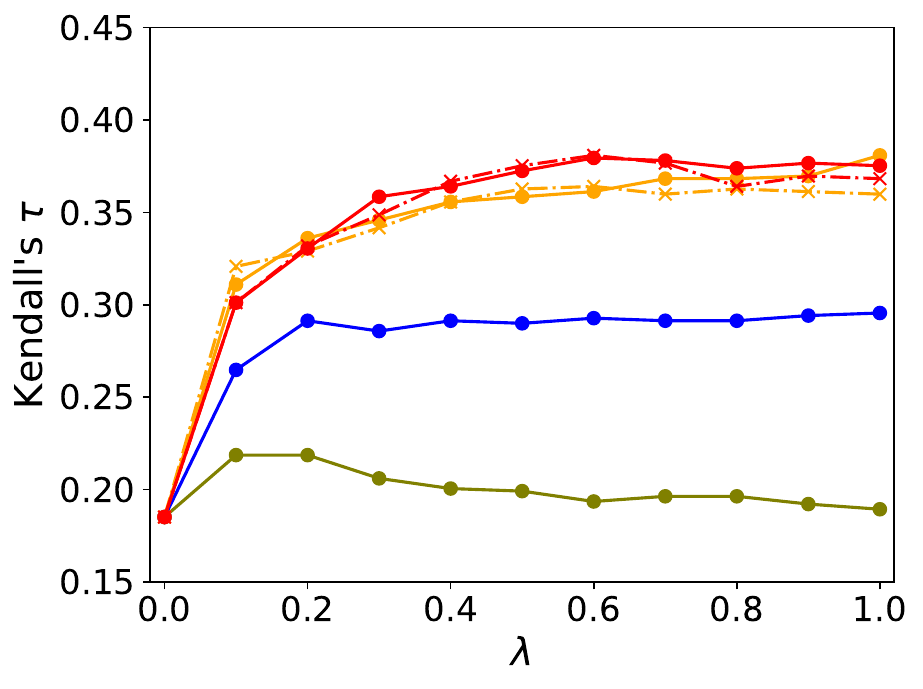}
         \caption{DL'20: $k=5$}
    \end{subfigure}
\caption{Effect of jointly varying both $k$ (the number of QVs leveraged in QPP) and $\lambda$ (relative importance of the QVs) on DL'19 (upper row) and DL'20 (lower row) test queries. The base estimator is NQC, and the target IR metric is AP@100.
In general, small values of $k$ perform best on both datasets. Whereas for $\lambda$, the optimal range is not consistent: compared with DL'19 queries, DL'20 queries require higher importance on QVs ($\lambda$) in the QPP estimations of Equation~\eqref{eq:generic-knn-qpp}.}
\label{fig:lambda_graphs}
\end{figure}

\para{Effect of varying both $k$ and $\lambda$ in QPP based on retrieved QVs}

Figure~\ref{fig:lambda_graphs} shows the joint effect of both the parameters $k$ and $\lambda$ on the retrieved-QV-based QPP models. These results are not the average of a 2-fold setup with hyperparameter tuning after train/test splits; instead, they show the parameter variation effects on TREC DL'19 and DL'20 test queries separately.

First, as a general trend, we observe from Figure~\ref{fig:lambda_graphs} that the retrieved QVs are mostly better than the baseline QPP methods based on W2V- and RLM-generated QVs. The only exception appears in TREC DL'19, where the QPP with the top-3 RLM-generated QVs produces better results than the proposed retrieved-QV-based QPP methods. In all other cases, the proposed methods perform better. Moreover, the QPP methods with 2-hop QVs are consistently better than the ones with 1-hop QVs, thus demonstrating the usefulness of the relevant documents in the training set.

To identify suitable values of the number of QVs ($k$), we examine the plots in Figure~\ref{fig:lambda_graphs} along the horizontal dimension (i.e., comparing subfigures~a, b, and c). For both the DL'19 and DL'20 query sets, the results indicate that smaller values of $k$ are more effective, with $k=1$ consistently yielding the best performance. However, the two datasets exhibit a notable discrepancy in the optimal value of $\lambda$. Specifically, the best-performing configurations for DL'19 tend to occur when $\lambda$ is close to 0.3, whereas for DL'20, the highest scores are concentrated on the right side of the $\lambda$-axis, where $\lambda>0.5$.

\begin{figure}[t]
\centering
\includegraphics[width=0.99\columnwidth]{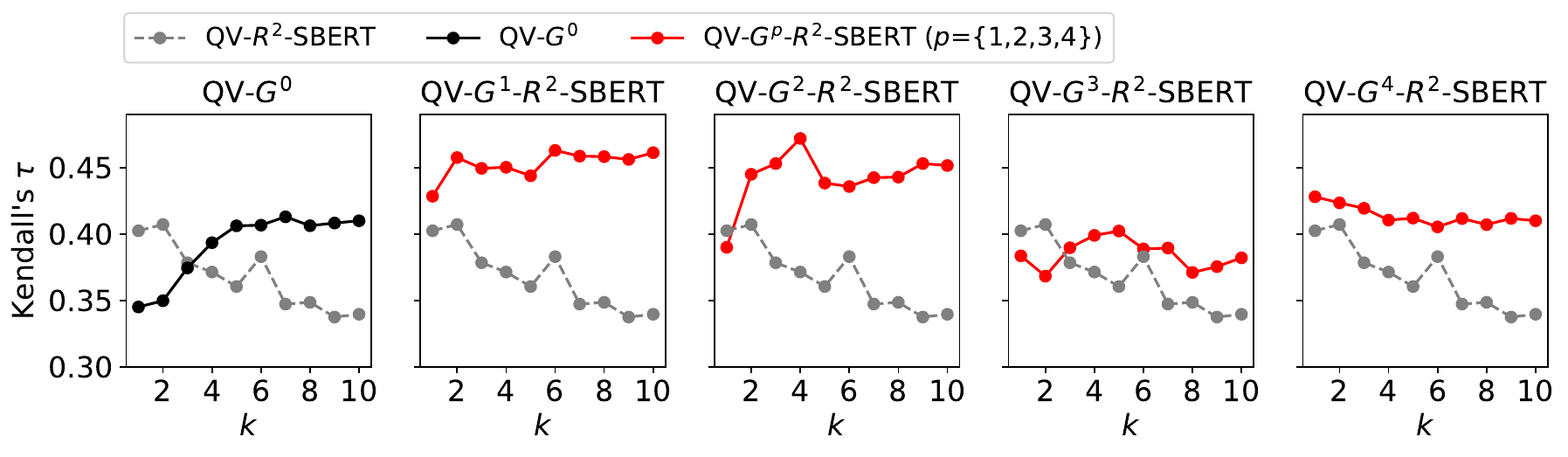}
\caption{Sensitivity of the RAQG-based QPP approaches on the context size $p$. From left to right, the number of retrieved QVs ($p$) in the RAQG context increments from $0$ to $4$. We take the retrieved QVs used in QV-$R^2$-SBERT -- the best performer in Table \ref{table:ret_qv_main} -- as the RAQG context. For ease of comparison, we show the QPP accuracy of QV-$R^2$-SBERT as grey points in each subplot. The y-axis ranges for all the sub-plots are identical.}

\label{fig:sbert_r-pshot}
\end{figure}

\subsubsection{Parameter Sensitivity in QPP Approaches with RAQG QVs}\label{sss:sensitivity-raqg}

In addition to the number of QVs ($k$) and the relative importance of the QVs ($\lambda$), there is a third hyperparameter for the proposed QPP approach based on Retrieval-Augmented QV Generation (RAQG): $p$ -- the number of retrieved QVs used as RAQG context. Taking the QVs used by QV-$R^2$-SBERT -- the best-performing retrieved-QV-based method in Table~\ref{table:ret_qv_main} -- as RAQG context, in this section, we analyse the parameter sensitivity of RAQG-based QPP with different $p$ values.

\para{Effect of varying $k$ with respect to $\lambda$ in RAQG-based QPP}
Figure~\ref{fig:sbert_r-pshot} illustrates how QPP accuracy varies with $k$, the number of retrieved QVs used for contextual generation, in RAQG-based QPP models. In contrast to the decreasing trend observed in retrieved-QV-based approaches -- where increasing $k$ typically leads to reduced QPP accuracy -- the accuracy of RAQG-based QPP generally improves as $k$ increases. This behaviour can be attributed to the generative mechanism’s ability to produce a large pool of QVs that consistently express a flexible but realistic underlying information need, a property that does not hold when QVs are retrieved from a finite training set of queries. While a small number of LLM-generated QVs may offer limited benefit, increasing $k$ allows the aggregated influence of a larger set of QVs to take effect, ultimately enhancing QPP performance. Notably, this upward trend along the $k$ axis is most pronounced for QV-$G^0$ and becomes less prominent as the number of examples increases.

\para{Effect of the number of examples ($p$) in RAQG-based QPP}
The subplots in Figure~\ref{fig:sbert_r-pshot} present the QPP accuracy achieved by RAQG-based QPP with a varying number of retrieved QVs incorporated as contextual examples. By comparing these subplots, we observe that the highest correlations occur when one or two retrieved QVs are used as context ($p=1$ and $p=2$). In the 0-shot setting ($p=0$), QV-$G^0$ is able to surpass the best-performing retrieval-only model, QV-$R^2$-SBERT, but only when a sufficiently large number of generated QVs are used. When more than two retrieved QVs are employed as RAQG context ($p>2$), although the QPP accuracy remains generally higher than QV-$R^2$-SBERT's, adding more examples to RAQG context seems to harm QPP's accuracy. This decline in QPP accuracy may be attributed to the unstable quality of retrieved QVs. As observed in Section~\ref{sss:sensitivity-ret-only}, QPP with retrieved QVs performs poorly at large $k$s, suggesting a degradation in the quality of RAQG context if the context size $p$ increases beyond two. Therefore, as $p$ increases, these low-quality retrieved example QVs may result in generating QVs that offer diminishing improvements in QV-based QPP.

\section{Discussion}\label{sec:discussions}

In this section, we discuss four questions related to the underlying mechanism of the proposed QV-based QPP framework. Section~\ref{ss:inner_retriever} examines the effect of separating the internal retriever and target retriever in Equation~\eqref{eq:generic-knn-qpp} through an ablation study. Section~\ref{ss:qv_effect_case_study} analyses how leveraging QVs enhances QPP accuracy for neural retrievers by comparing the distributions of QPP estimation. Section~\ref{ss:qv_characteristics} presents a case study in which we inspect the specific QVs generated by RAQG and discuss the characteristics of the QVs that positively impact QPP accuracy. Section~\ref{ss:out-of-distribution} evaluates the proposed QPP framework in an out-of-distribution setting, where target queries retrieve documents from a corpus with different statistical characteristics from those used in the main experiments.

\subsection{Does employing a neural internal retriever enhance QPP accuracy?}\label{ss:inner_retriever}

\begin{table}[t]
\centering
\caption{Effectiveness of QV-based QPP when the internal retriever $M'$ is identical to the target retriever $M$. It is an ablation study for the left two columns in Table~\ref{table:ret_qv_main}, where BM25$\gg$MonoT5 is the target retriever. Except for using BM25$\gg$MonoT5 as $M'$, the other settings of the ablation study remain the same as the main experiments.
The best $\tau$ values for predicting each target metric (AP@100 and nDCG@10) are bold-faced. Compared with the results obtained with $M'=BM25$, the QPP effectiveness decreases significantly (a $\dagger$ alongside a reported number indicates the decrease is within the 90\% confidence level in Fisher's $z$ test).}
\begin{adjustbox}{width=0.55\columnwidth}    
\begin{tabular}{@{}ll ll ll@{}}
\toprule

\multicolumn{2}{c}{$M=M'=\text{BM25$\gg$MonoT5}$} & \multicolumn{2}{c}{$\phi=\text{NQC}$}&
\multicolumn{2}{c}{$\phi=\text{UEF}$} \\
\cmidrule(r){3-4} \cmidrule(r){5-6}
Type & Method & AP-$\tau$  & nDCG-$\tau$   & AP-$\tau$  & nDCG-$\tau$\\
\midrule 
\multirow{3}{*}{\rotatebox[origin=c]{0}{BL}} 
   & $\phi$ & 0.1673 & 0.0274 & 0.1679 & 0.0341 \\
& QV-W2V & 0.2035 & \textbf{0.1389} & 0.2123 & \textbf{0.1287} \\
& QV-RLM & 0.1728$\dagger$ & -0.0020$\dagger$ & 0.1414$\dagger$ & 0.0264$\dagger$ \\
\cmidrule(r){2-6}

 \multirow{4}{*}{\rotatebox[origin=c]{0}{Ours}} 
   & QV-$R^1$-BM25 & \textbf{0.2676} & 0.0963 & \textbf{0.2772} & 0.0495$\dagger$ \\
 & QV-$R^1$-SBERT & 0.2583 & 0.0723$\dagger$ & 0.2473 & 0.0481$\dagger$ \\ 
 & QV-$R^2$-BM25 & \textbf{0.2676}$\dagger$ & 0.0972$\dagger$ & 0.2754 & 0.0474$\dagger$ \\ 
 & QV-$R^2$-SBERT & 0.2527$\dagger$ & 0.0635$\dagger$ & 0.2480$\dagger$ & 0.0285$\dagger$ \\

\bottomrule
\end{tabular}
\end{adjustbox}

\label{table:neural_inner_ret}
\end{table}

In addition to the QV retrieval and Retrieval-Augmented QV Generation (RAQG) component, our proposed methodology introduces an independent internal retriever as shown in the upper-right part of Figure~\ref{fig:flowchart}. As noted after Equation~\eqref{eq:generic-knn-qpp}, in our experiments for QPP on neural rankers, we use BM25 as the internal retriever $M'$, which is agnostic of the target retrieval pipeline $M$. Table~\ref{table:neural_inner_ret} presents the QPP accuracy when $M\equiv M'\gets \text{BM25$\gg$MonoT5}$. These results can be compared with the results reported in Table~\ref{table:ret_qv_main}, where BM25 was used as $M'$.

By comparing Table~\ref{table:neural_inner_ret} with the left two columns (about BM25$\gg$MonoT5) in Table~\ref{table:ret_qv_main}, we
observe a consistent decrease in the correlation values of every tested QV-based QPP method, including the baselines QV-W2V and QV-RLM. For instance, QV-RLM AP correlation drops from $0.3308$ to $0.1728$, whereas the QV-$R^2$-SBERT correlation drops from $0.4033$ to $0.2527$. Similarly, when it comes to predicting nDCG@10, the gains achieved with the use of QVs decrease even more substantially.

From the above comparisons, we conclude that when the target retriever involves a neural ranker, employing the same ranker as the internal retriever for QV-based QPP, as formulated in Equation~\eqref{eq:generic-knn-qpp}, does not improve the prediction accuracy. This may be attributed to the base predictor's (NQC and UEF in our experiments) limited accuracy for neural rankers~\cite{WRIG}. Decoupling the target and the internal retrievers provides an advantage to our QPP setup in the sense that our framework can operate with an internal retriever that is not only efficient (sparse model) but is also agnostic of the target retriever.

\subsection{How does leveraging the QVs generated by the proposed RAQG enhance QPP accuracy?}\label{ss:qv_effect_case_study}

\begin{figure}
\centering
\begin{adjustbox}{width=0.8\textwidth}

\begin{subfigure}{0.4\textwidth}
 \centering
 \includegraphics[width=\textwidth]{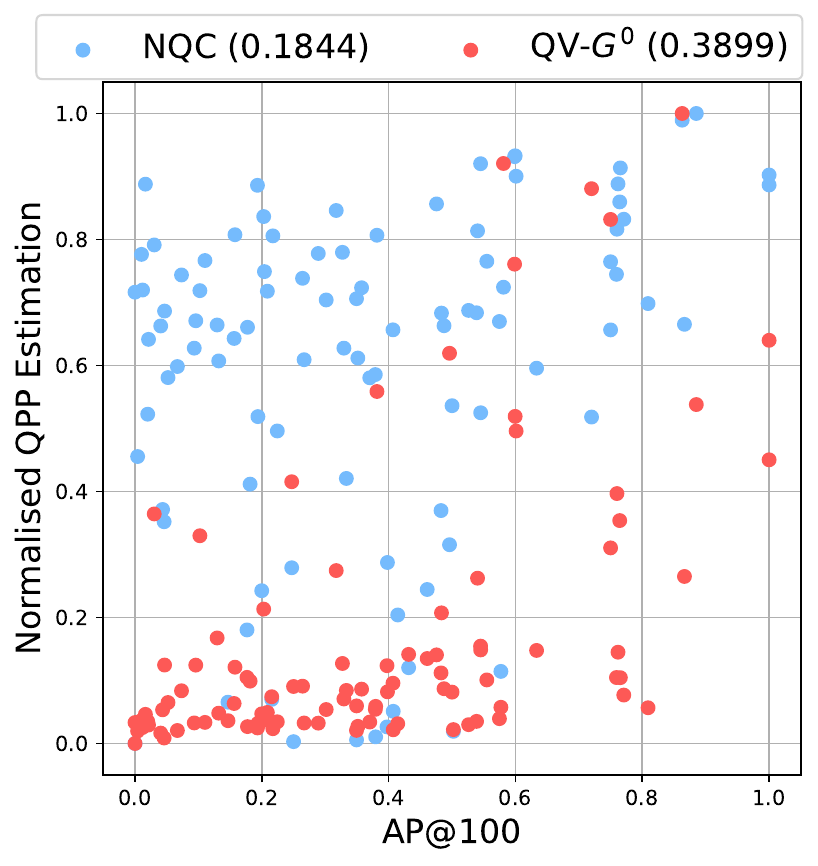}
 \caption{NQC vs QV-$G^0$}
 \label{fig:apuef}
\end{subfigure}
\hfill
\begin{subfigure}{0.4\textwidth}
 \centering
 \includegraphics[width=\textwidth]{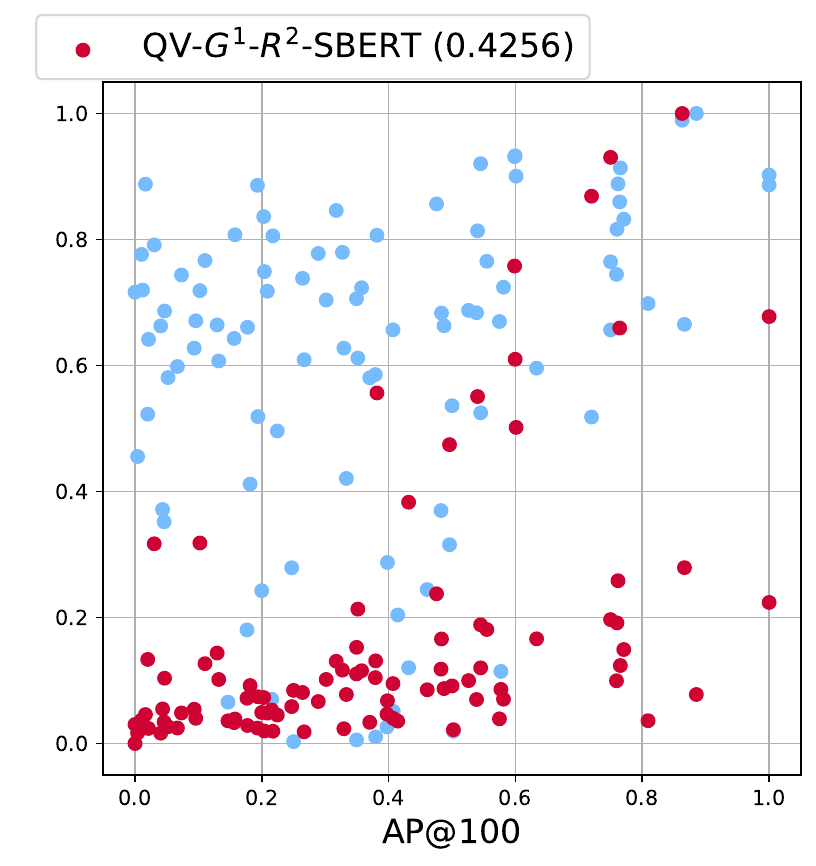}
 \caption{NQC vs QV-$G^1$-$R^2$-SBERT}
 \label{fig:ndcgnqc}
\end{subfigure}
\end{adjustbox}
\caption{Scatter plots of the normalised QPP estimation vs AP@100 values of the 97 queries in TREC DL'19 and DL'20 test sets, the base predictor is NQC. The reported numbers are about the QPP methods (a) QV-$G^0$, which is the best baseline without RAQG, and (b) QV-$G^1$-$R^2$-SBERT, which is the best proposed QPP method. Only one QV is applied for both methods ($k=1$). We fix $\lambda=0.9$, which is the optimal setting for both tested methods at $k=1$. NQC achieves Kendall's $\tau$ at $0.1844$, while the QPP methods of QV-$G^0$ and QV-$G^1$-$R^2$-SBERT achieve $0.3621$ and $0.4256$, respectively.}
\label{fig:qv_effects_lam=0.9}

\end{figure}

So far, we have reported correlations between target IR metrics and QPP estimates. We now provide a per-query visual analysis using scatter plots between AP@100 and the corresponding QPP estimates for queries from the combined TREC DL'19 and DL'20 sets. To avoid clutter, we compare the best-performing QV-based method, QV-$G^1$-$R^2$-SBERT, with its 0-shot counterpart QV-$G^0$ and the base predictor NQC.

Figure~\ref{fig:qv_effects_lam=0.9} shows the normalised QPP estimates produced by NQC, QV-$G^0$, and RAQG-generated QVs. The x-axis represents AP@100, while the y-axis represents the corresponding QPP estimate. As shown in Figure~\ref{fig:qv_effects_lam=0.9}(a), leveraging 0-shot generated QVs already improves over the base predictor by reducing the number of queries with low AP@100 but high predicted QPP scores (i.e., outliers in the upper-left region). This suggests that incorporating QVs helps suppress overly optimistic predictions for poorly performing queries from the base predictor. A comparison between Figure~\ref{fig:qv_effects_lam=0.9} (a) and (b) reveals that RAQG QVs further reduce such outliers at the low-AP end. Moreover, RAQG also reduces erroneous suppression for some high-AP queries (i.e., fewer points in the bottom-right region). These differences, although subtle visually, result in a noticeably higher Kendall’s $\tau$ correlation.

These observations suggest that RAQG QVs, which are generated by grounding the LLM in retrieved real-user queries, help mitigate extreme mispredictions for low-performance queries while better retaining signals for high-performance ones when incorporated into QPP.

\subsection{What characteristics of query variants (QVs) enhance QPP accuracy?}\label{ss:qv_characteristics}

\begin{figure}
\centering
    \begin{adjustbox}{width=0.95\textwidth}
    \includegraphics{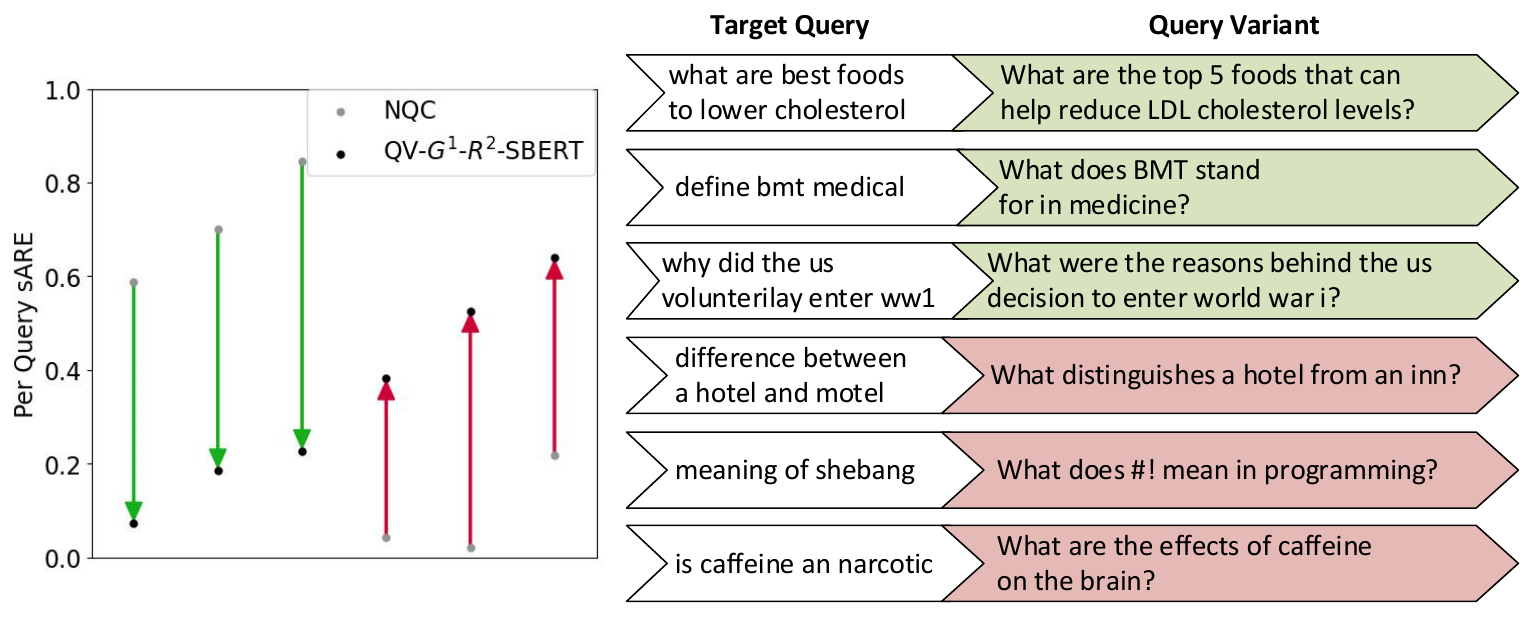}
    \end{adjustbox}
\caption{A case study of the queries where per-query sARE values are reduced or increased the most compared to the base predictor NQC. The results are obtained using QV-$G^1$-$R^2$-SBERT, the best-performing proposed QPP method, where only one QV is used ($k$=1). A decrease in per-query sARE for a target query indicates that leveraging the top-1 QV in QPP estimation makes its ranking in the test query set (97 queries from TREC DL'19 and DL'20 combined) more aligned with its true ranking by AP@100. The green arrows indicate decreases in sARE, while the red arrows indicate increases in sARE. The RAQG QVs used in QPP for the target queries are listed on the right side of this figure, matching the order of queries in the plot (from left to right).
}
\label{fig:case_study}
\end{figure}

After analysing how leveraging QVs adjusts the QPP estimations over a query set, we now conduct a case study of individual target queries to identify common characteristics of QVs that positively or negatively affect QPP estimation. Since absolute QPP estimations are affected by changes in scale when leveraging QVs, we use per-query Scaled Absolute Relative Error (sARE)~\citep{SMARE} to assess whether a QV improves or worsens the prediction quality for a target query. A lower per-query sARE than that of the base predictor indicates that the QV makes the predicted ranking of the query more consistent with its true retrieval performance ranking. Following the setting in Section~\ref{ss:qv_effect_case_study}, we conduct this case study using QV-$G^1$-$R^2$-SBERT (the best-performing proposed QPP method).

Figure~\ref{fig:case_study} shows the target queries with the largest decreases and increases in sARE compared with the base predictor. For queries where sARE decreases (QVs in green), the QVs tend to more faithfully capture and clarify the underlying information need. For instance, in the query concerning the First World War, the generated QV introduces terms such as ``reason'' and ``decision'', which help refine and make the intent more explicit. In contrast, QVs associated with performance degradation (QVs in red) often drift from the original topic or omit essential concepts. For example, for the query ``is caffeine a narcotic,'' the generated QV fails to include the term ``narcotic'' and instead produces a broader information need focused on the neurological effects of caffeine. Even in cases where the QV appears to preserve the query's intent---such as when the term ``shebang'' is replaced by its programmatic form ``\#!''---the resulting variant may hinder retrieval effectiveness, as most relevant documents contain the lexical form rather than the symbolic representation.

Beyond qualitative inspection, we also observe that a QV's usefulness in QPP is not determined solely by RBO-based topical similarity. Although RAQG-generated QVs do not always maximise Rank-Biased Overlap (RBO) with the target query, they are often formulated more concisely, which is similar to the target user queries. This observation suggests that strict topical similarity is not the sole determinant of QV usefulness for QPP, factors such as the query formulation may also matter.

\begin{table}[t]
\centering
\caption{Results of applying the proposed QV-based QPP methods on TREC DL'21 and DL'22 test sets. These datasets are constructed over a different passage corpus from the MS MARCO passage collection used in the main experiments, providing an out-of-domain evaluation. Query variants are retrieved from the MS MARCO training query set. The target retrieval pipeline is BM25$\gg$MonoT5, and the base QPP predictor is NQC. Results are reported with $\lambda = 0.9$, which is the optimal setting identified in prior experiments. For each value of $k$ (the number of query variants leveraged in QPP), the best result in each column is underlined; the overall best result in each column is shown in bold.}
\begin{adjustbox}{width=0.6\columnwidth} 
\small
\begin{tabular}{@{}ll cc ll ll @{}}
\toprule
& & & & \multicolumn{2}{c}{TREC-DL-21} & \multicolumn{2}{c}{$\text{TREC-DL-22}$} \\
\cmidrule(r){5-6} \cmidrule(r){7-8} 
$k$ & Method & R & G & AP-$\tau$ & nDCG-$\tau$ & AP-$\tau$ & nDCG-$\tau$  \\
\midrule
0 & NQC &  &  & -0.0160 & -0.0400 & 0.1474 & 0.1483 \\
\cmidrule{1-8}
\multirow{4}{*}{1} & QV-$R^2$-BM25 & \cmark &  & 0.1237 & 0.1547 & 0.1700 & 0.0482 \\
& QV-$R^2$-SBERT & \cmark &  & 0.0596 & -0.0618 & 0.1199 & -0.0081 \\
\cmidrule{3-8}
& QV-$G^1$-$R^2$-BM25 & \cmark & \cmark & \textbf{\underline{0.2865}} & \textbf{\underline{0.2230}} & 0.1777 & 0.1053 \\
& QV-$G^1$-$R^2$-SBERT & \cmark & \cmark & 0.2415 & 0.1954 & \textbf{\underline{0.2856}} & \textbf{\underline{0.1807}} \\
\cmidrule{1-8}
\multirow{4}{*}{5} & QV-$R^2$-BM25 & \cmark &  & 0.0829 & 0.0937 & 0.1594 & 0.0814 \\
& QV-$R^2$-SBERT & \cmark &  & 0.0815 & 0.0400 & 0.1446 & 0.0694 \\
\cmidrule{3-8}
& QV-$G^1$-$R^2$-BM25 & \cmark & \cmark & \underline{0.1978} & \underline{0.1998} & 0.1382 & 0.0990 \\
& QV-$G^1$-$R^2$-SBERT & \cmark & \cmark & 0.1789 & 0.1751 & \underline{0.2363} & \underline{0.1574} \\

\bottomrule
\end{tabular}

\label{table:dl_21_22}
\end{adjustbox}
\end{table}

\subsection{How well does the proposed QV-based QPP framework generalise to out-of-distribution datasets?}\label{ss:out-of-distribution}

To evaluate the generalisability of our proposed QPP methods beyond the TREC DL'19 and DL'20 datasets, we further evaluate on TREC DL’21~\cite{TREC-DL-2021-overview} and TREC DL’22~\cite{TREC-DL-2022-overview}, which were created over the MS MARCO passage v2 corpus. Although the two corpora are closely related, they differ in their data distributions and retrieval characteristics. This constitutes an out-of-distribution evaluation setting, allowing us to assess how well our QPP methods perform when the underlying retrieval corpus and query–document relationships differ from those in the original benchmarks. Same as the main experiments, in this experiment, we retrieve query variants from the MS MARCO training query set.

Table~\ref{table:dl_21_22} reports the prediction accuracy of our proposed QPP methods leveraging 2-hop retrieved QVs and contextually generated RAQG QVs on TREC DL’21 and DL’22. Overall, both retrieved-QV-based and RAQG-based QPP methods improve over the base predictor, with the best Kendall’s $\tau$ values for AP@100 reaching around 0.3 and for nDCG@10 around 0.2. These results indicate that the effectiveness of query variants for QPP extends beyond the original evaluation on TREC DL’19 and DL’20.

Compared with the main observations on MS MARCO passage benchmarks, two notable differences emerge in this out-of-distribution setting. First, the gap between QPP performance using retrieved QVs alone and using RAQG QVs is generally larger. One possible explanation is that the TREC DL’21 and DL’22 queries are less likely to find closely related variants in the MS MARCO training query set. In such cases, RAQG can generate query variants that better match the underlying information need and better resemble real user queries, leading to larger gains in accuracy.
Second, increasing the number of query variants ($k$) does not improve prediction accuracy (as observed from Table~\ref{table:pshot-k=1}) on these datasets. In contrast to the main experiments, the best results are often obtained with a small number of QVs (e.g., $k = 1$), suggesting that simply adding more variants can introduce noise rather than useful signal in the out-of-distribution context.
Together, these findings suggest that when deploying QV-based QPP methods on a target corpus with different retrieval characteristics, it is important to tune key configuration parameters such as the number of query variants and the balance between retrieved versus generated QVs.

\section{Conclusions}\label{sec:conc}

In this work, we introduced a retrieval-oriented perspective on query variants (QVs) for query performance prediction (QPP), departing from prior approaches that rely exclusively on QVs generated by relevance models or non-contextual embeddings. We proposed a method for retrieving QVs from a query collection, expanding the candidate space by leveraging the relevant documents of 1-hop neighbours to surface additional queries that may encode the same underlying information need. These retrieved QVs can be used directly within QV-based QPP or employed as contextual examples to guide large language models in generating new QVs through a framework we term Retrieval-Augmented QV Generation (RAQG).

Our study demonstrates that retrieved QVs and RAQG offer a flexible and effective foundation for improving QPP. The framework accommodates both low-cost QV retrieval and LLM-assisted QV generation, enabling practitioners to balance computational constraints against performance goals. This adaptability is particularly valuable in scenarios where query logs are available and can be exploited to derive richer query variants.

More broadly, RAQG provides a principled mechanism for generating QVs that support downstream information retrieval tasks beyond QPP, including query reformulation and retrieval-augmented generation. An important direction for future work is to improve the controllability and utility of RAQG. Rather than supplying a fixed set of retrieved examples, dynamic example-selection strategies may enable more targeted and effective QV generation. Exploring such mechanisms may offer a promising path toward producing high-quality QVs that further enhance QPP and other retrieval-driven applications.

\bibliographystyle{elsarticle-num-names} 

\end{document}